\documentclass[reprint,nofootinbib,superscriptaddress,amsmath,amssymb,aps,prl]{revtex4-1}

\usepackage{graphicx}
\usepackage{dcolumn}
\usepackage{bm}
\usepackage{todonotes}
\usepackage{simplewick}
\usepackage{hyperref}
\hypersetup{
	colorlinks,
	linkcolor={blue!50!black},
	citecolor={blue!50!black},
	urlcolor={blue!80!black}
}
\graphicspath{{./figures/}}
\usepackage{pgffor}
\usepackage{xspace}
\makeatletter
\DeclareRobustCommand\onedot{\futurelet\@let@token\@onedot}
\def\@onedot{\ifx\@let@token.\else.\null\fi\xspace}

\makeatother

\begin{document}
	
	\title{Ab initio limits of atomic nuclei}
	
	\author{S.~R.~Stroberg}
	\email[Email:~]{stroberg@uw.edu}
	\affiliation{Department of Physics, University of Washington, Seattle, WA 98195, USA}
	\affiliation{TRIUMF, 4004 Wesbrook Mall, Vancouver, BC V6T 2A3, Canada}
	
	\author{J.~D.~Holt}
	\email[Email:~]{jholt@triumf.ca}
	\affiliation{TRIUMF, 4004 Wesbrook Mall, Vancouver, BC V6T 2A3, Canada}
	\affiliation{Department of Physics, McGill University, 3600 Rue University, Montr\'eal, QC H3A 2T8, Canada}
	
	\author{A.~Schwenk}
	\email[Email:~]{schwenk@physik.tu-darmstadt.de}
	\affiliation{Institut f\"ur Kernphysik, Technische Universit\"at Darmstadt, 64289 Darmstadt, Germany}
	\affiliation{ExtreMe Matter Institute EMMI, GSI Helmholtzzentrum f\"ur Schwerionenforschung GmbH, 64291 Darmstadt, Germany}
	\affiliation{Max-Planck-Institut f\"ur Kernphysik, Saupfercheckweg 1, 69117 Heidelberg, Germany}
	
	\author{J.~Simonis}
	\email[Email:~]{simonis@uni-mainz.de}
	\affiliation{\mbox{Institut f\"ur Kernphysik and PRISMA Cluster of Excellence, Johannes Gutenberg-Universit\"at, 55099 Mainz, Germany}}
	\affiliation{Institut f\"ur Kernphysik, Technische Universit\"at Darmstadt, 64289 Darmstadt, Germany}
	\affiliation{ExtreMe Matter Institute EMMI, GSI Helmholtzzentrum f\"ur Schwerionenforschung GmbH, 64291 Darmstadt, Germany}
	
	\begin{abstract}
		We predict the limits of existence of atomic nuclei, the proton and neutron drip lines, from the light through medium-mass regions.
		Starting from a chiral two- and three-nucleon interaction with good saturation properties, we use the valence-space in-medium similarity renormalization group to calculate ground-state and separation energies
		from helium to iron, nearly 700 isotopes in total.
		We use the available experimental data to quantify the theoretical uncertainties for our ab initio calculations towards the drip lines.
		Where the drip lines are known experimentally, our predictions are consistent within the estimated uncertainty.
		For the neutron-rich sodium to chromium isotopes, we provide predictions to be tested at rare-isotope beam facilities.
	\end{abstract}
	
	\maketitle
	
	Atomic nuclei, which form the basis for known matter in the Universe, cannot be made from arbitrary numbers of protons and neutrons. 
	For a given element (i.e., proton number $Z$) a nucleus can support only so many neutrons, $N$, and vice versa. 
	The point at which nucleons no longer form a bound system is referred to as the drip line.
	Specifically, at the drip line one- or two-nucleon separation energies become negative, and nuclei decay via nucleon emission. 
	The proton drip line is known experimentally to the medium-mass region, but to date, the neutron drip line is established only up to neon ($Z=10$)~\cite{Thoe16Book,Ahn2019}. 
	Pinning down the neutron drip line to calcium and beyond is a flagship scientific motivation for next-generation rare-isotope beam facilities~\cite{Gade16FRIB,Dura19FAIR}. 
	Indeed several neutron-rich isotopes, including $^{60}$Ca, were recently discovered in this region~\cite{Tara18Ca60}. 
	Furthermore, knowledge of the neutron drip line is important
	for $r$-process simulations modeling the synthesis of heavy elements~\cite{Mump15PPNP,Mart16massrpro} that occurs in neutron-star mergers~\cite{Abbo17mm}.
	
	Predicting the location of the drip lines
	poses a substantial theoretical challenge, particularly because many nuclei far from known data must be calculated systematically. 
	In a pioneering study, the nuclear landscape was predicted from extrapolations of state-of-the-art nuclear density functional theory, and approximately 7000 nuclei were estimated to exist in nature~\cite{Erle12Nature}. 
	Since this work, tremendous progress has been made in statistical analyses of nuclear models~\cite{Neuf2019,Yoshida18} as well as in ab initio nuclear theory. 
	Developments in chiral effective field theory~\cite{Epel09RMP,Mach11PR,Hamm13RMP} and similarity renormalization group~\cite{Bogn07SRG,Bogn10PPNP} are pushing nuclear forces to new levels of accuracy and ranges of applicability.
	Though a robust and systematic theoretical framework has not yet been fully achieved, particular nuclear Hamiltonians have been constructed which reproduce ground-state energies up to the tin region~\cite{Hebe11fits,Ekst15sat,Morr17Tin}.
	Three-nucleon (3N) forces play a key role
	for understanding the drip lines~\cite{Otsu10Ox,Hage12Ox3N,Cipo13Ox,Herg13MR,Holt13PR,Hebe15ARNPS}.
	Moreover, many-body theories~\cite{Barr13PPNP,Hage14RPP,Carl15RMP,Herg16PR,Stroberg19ARNPS,Barb17SCGFlectnote} have advanced to treat medium-mass open-shell systems~\cite{Herg13MR,Bogn14SM,Jans14SM,Soma14GGF2N3N,Stro17ENO}, with the primary limitation being computational resources needed to obtain convergence with respect to basis size, laying the groundwork for a new era of ab initio theory.
	
	In this Letter we calculate properties of essentially all nuclei from helium to iron ($Z = 2 - 26$), close to 700 in total, to provide a global ab initio survey of ground-state energies and predict the nuclear drip lines.
	Using two-nucleon (NN) and 3N interactions constrained by only few-body data, we solve the many-body problem with the valence-space formulation of the in-medium similarity renormalization group (VS-IMSRG)~\cite{Tsuk12SM,Herg16PR,Bogn14SM,Stro16TNO,Stro17ENO,Stroberg19ARNPS}.
	Our results yield an overall root-mean-square (rms) deviation of 3.3~MeV from absolute experimental energies and 0.7-1.4~MeV from separation energies.
	In comparison, state-of-the-art energy-density functionals obtain rms devations in 
	the range 0.6--2.0~MeV for energies and 0.4--1.25~MeV for separation energies~\cite{Goriely2009,Bulgac2018,NavarroPerez2018,Neuf2018} (note, however, that the density functional rms values are obtained over a much larger range of masses). 
	
	\begin{figure*}[t]
		\includegraphics[width=\textwidth]{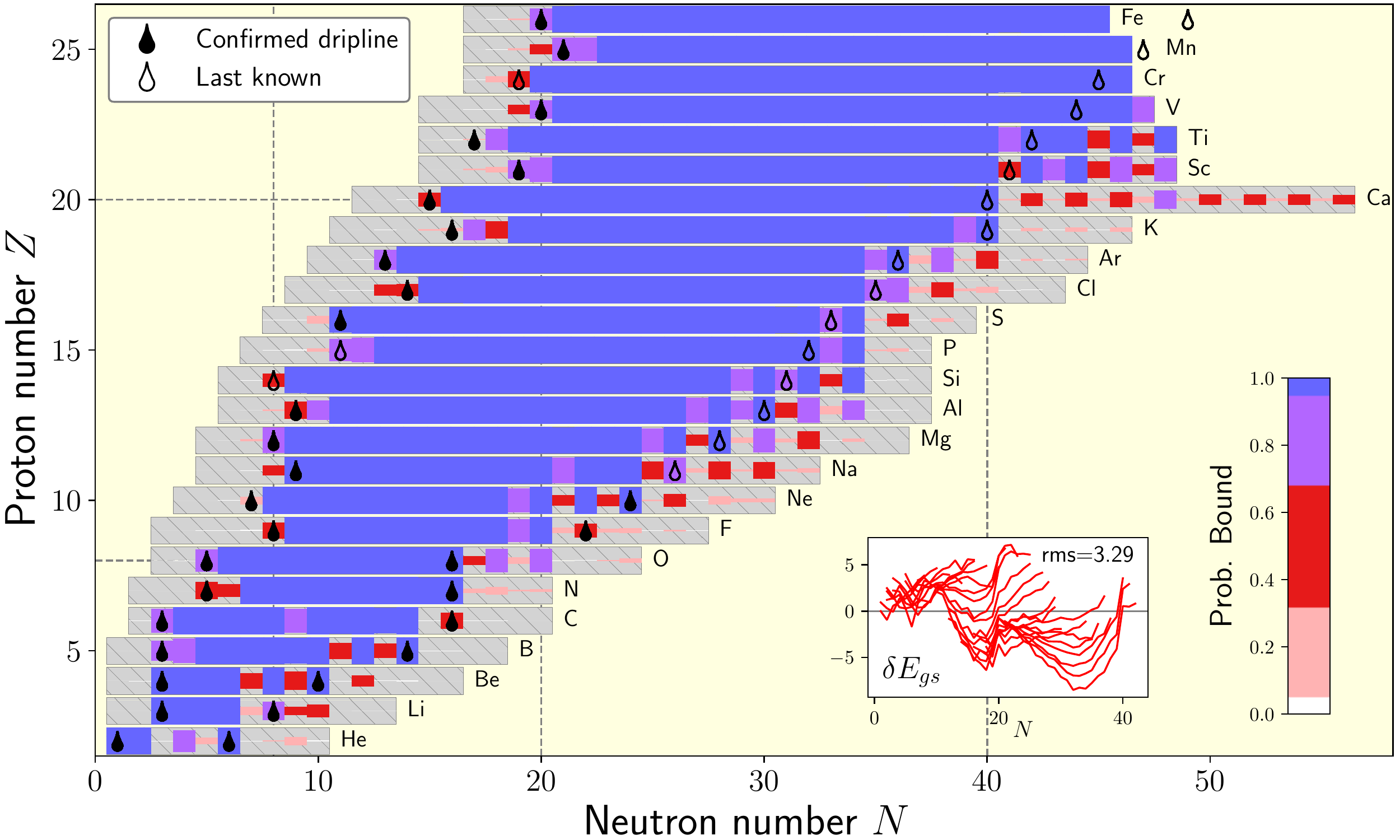}
		\caption{Calculated probabilities for given isotopes to be bound with respect to one- or two-neutron/proton removal. The gray region indicates nuclei that have been calculated, while the height of the boxes corresponds to the estimated probability that a given nucleus is bound with respect to one- or two-neutron (proton) removal in the neutron-rich (deficient) region of the chart. The inset shows the residuals with experimental ground-state energies.}
		\label{drip_unc}
	\end{figure*}
	
	While the drip line signature is unambiguous experimentally, from a theoretical perspective an error of a few tens of keV -- well beyond current levels of precision -- can make the difference between an isotope being bound or unbound.
	Therefore, an assessment of theoretical uncertainty is mandatory for any meaningful drip line prediction.
	Ab initio methods present an appealing framework for uncertainty quantification: one begins with the most general Lagrangian compatible with the applicable symmetries, organized by a systematically improvable power counting, then solves the nuclear many-body problem within a controlled and systematically improvable approximation scheme, propagating all uncertainties.
	Such a prescription has not yet been achieved in practice, so for the present we use a comparison with known data to calibrate a physically motivated model for the error.
	Recent work in a similar spirit has applied Bayesian machine learning algorithms to global mass models~\cite{Neuf2018,Neuf2019,Neuf2020}.
	The main advantages of our current approach are (i) the predictions should not be biased towards measured data, because they were not fit to any data beyond helium and (ii) the predictions can be benchmarked where the proton and neutron drip lines are known experimentally (mass models are typically applied to $Z \gtrsim 8$). 
	
	In the VS-IMSRG, a valence-space Hamiltonian of tractable dimension is decoupled from the larger Hilbert space via an approximate unitary transformation. 
	We begin in a harmonic-oscillator basis of 15 major shells (i.e., $e=2n+l \leqslant e_{\mathrm{max}}=14$) with an imposed cut of $e_{1}+e_{2}+e_{3} \leqslant E_{\mathrm{3Max}}=16$ for 3N matrix elements. The resulting ground-state energies are converged to better than a few hundred keV with respect to these truncations, and we perform extrapolations in $e_{\rm max}$ to obtain infrared convergence~\cite{Furnstahl2012,Furnstahl2015}.
	Transforming to the Hartree-Fock basis, we capture effects of 3N interactions between valence nucleons via the ensemble normal-ordering of Ref.~\cite{Stro17ENO}.
	We then use the Magnus formulation of the IMSRG~\cite{Morr15Magnus,Herg16PR}, truncating all operators at the normal-ordered two-body level---the IMSRG(2) approximation---to generate approximate unitary transformations that decouple the core energy and valence-space Hamiltonian for each nucleus to be calculated.
	
	By default, we employ a so-called $0\hbar\omega$ valence space, where valence nucleons occupy the appropriate single major harmonic-oscillator shell (e.g., for $8 < N(Z) < 20$ the $sd$ shell, $20 < N(Z) < 40$ the $pf$ shell, etc.). At $N(Z)=2,8,20,40$,
	we do not decouple a neutron (proton) valence space, and no explicit neutron (proton) excitations are allowed in the calculation.
	We discuss exceptions to this below.
	Finally the resulting valence-space Hamiltonians are diagonalized with the NuShellX@MSU shell-model code~\cite{Brow14NuShellX}
	(with the exception of a few of the heaviest Ca, Sc and Ti isotopes, which were computed with the m-scheme code K-shell~\cite{Shimizu2019}).
	
	We thus calculate ground (and excited) states of all nuclei from helium to iron, except those for which the shell-model diagonalization is beyond our computational limits.
	For the input NN+3N interaction, we use the 1.8/2.0\,(EM) potential of Refs.~\cite{Hebe11fits,Simo16unc}, where the 3N couplings were fit to the $^3$H binding energy and the $^4$He charge radius.
	This interaction reproduces experimental ground-state energies of light- to medium-mass nuclei remarkably well~\cite{Simo17SatFinNuc,Morr17Tin}.
	Studies of nuclear matter~\cite{Hebe11fits,Dris19nmat} have shown that this interaction saturates with slightly too much binding and at at somewhat too high density, leading to too small radii for finite nuclei~\cite{Simo17SatFinNuc}.
	We use this observation below to model our systematic error in the separation energies.
	In the Supplemental Material, we provide results for absolute and separation energies for all nuclei calculated.
	In the inset of Fig.~\ref{drip_unc}, we plot the range of agreement with experiment and find an overall rms deviation of 3.3~MeV.
	The experimental binding and separation energies are taken from the Atomic Mass Evaluation~\cite{Wang2017}, with additional recent data from Refs.~\cite{Webb2019,Leblond2018,Mukha2018,Michimasa2018}. 
	
	As no experimental input beyond $^4$He is used in the current calculations, our results should not be biased toward known data\footnote{One might argue that the selection of one interaction from the family of interactions used in~\cite{Simo17SatFinNuc} constitutes incorporation of information beyond $A=4$. However, in the cases we have checked, other interactions in the family yield similar separation energies.}.
	Therefore our approach is to use measured data to assess how well separation energies are reproduced in general, then assume our calculations will behave similarly for separation energies which have not yet been measured. 
	This neglects phenomena which may emerge in the neutron-rich region, such as halo structures or island-of-inversion physics~\cite{Warburton1990a,Caur14n20n28}, but our results suggest that the impact of these effects on separation energies tends to lie within our estimated uncertainties.
	
	To characterize the quality of the reproduction of experimental data, we assess the residual, $\delta S_{\alpha} = S_{\alpha}^{\rm{th}}-S_{\alpha}^{\rm{exp}}$, for the separation energy in channel  $\alpha \in\{n,p,2n,2p \}$.
	We model the residual as
	\begin{equation}
	\delta S = f(N,Z,S^{\rm th},\ldots) + \epsilon(\sigma^2) \,,
	\end{equation}
	where $f$ is a function characterizing the systematic error (see below) and $\epsilon$ is a random number drawn from a Gaussian distribution of mean 0 and variance $\sigma^2$.
	
	The main source of many-body error in our calculations is due to the IMSRG(2) approximation.
	For soft input interactions such as the one used here, this approximation is accurate for binding energies at the level of a few percent~\cite{Stro17ENO}.
	Given that binding energies in this region are a few hundred MeV, this would naively suggest an error of several MeV on the separation energies.
	However, the errors made for neighboring nuclei are strongly correlated (see Supplementary Material) and largely cancel, improving the accuracy of the separation energies (the uncorrelated part will contribute to the random error $\epsilon$).
	This correlation is deteriorated in the case where a different valence space is used for the two binding energies entering into the separation energy.
	Wherever possible, we use a consistent valence space to compute separation energies, but this is not always possible.
	For example, to compute the $S_{2n}$ of $^{38}$Cl, we require the ground-state energy of $^{38}$Cl ($N$=21), which has a valence neutron in the $pf$ shell, and $^{36}$Cl ($N$=19), which has a valence hole in the $sd$ shell.
	This leads to an increased error for these special cases, and we treat these separately.
	
	\begin{figure}[t]
		\includegraphics[width=\columnwidth]{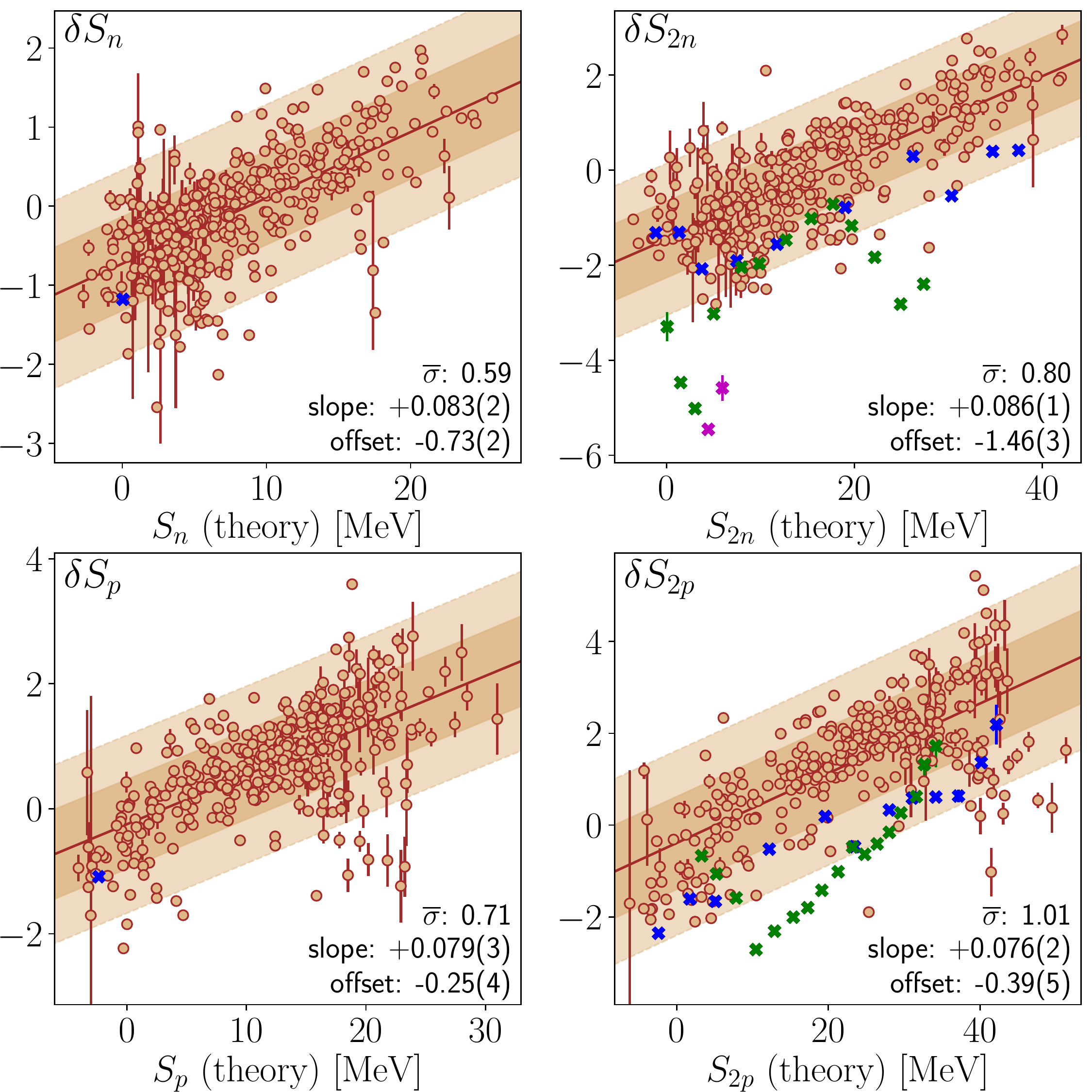}
		\caption{Dependence of the residuals $\delta S= S^{\rm th}-S^{\rm exp}$ on the computed separation energy. The brown line and bands reflect the mode (maximum of the distribution) and 68\% and 95\% confidence intervals of the posterior predictive distribution obtained by a Bayesian linear regression.  In each panel, the slope, offset and average standard deviation $\bar{\sigma}$ of the posterior predictive distribution are listed (in MeV). The crosses indicate cases in which the separation energy was computed with inconsistent valence spaces, due to the $N,Z= 8,20,40$ shell gaps indicated in blue, green and magenta, respectively.}
		\label{fig:Regression}
	\end{figure}
	
	We must also consider errors due to deficiencies of the input Hamiltonian.
	As mentioned above, a notable deficiency of this Hamiltonian is that predicted radii are systematically smaller than experiment by $\approx 4$\% for the mass range considered~\cite{Simo17SatFinNuc}.
	This should have an impact on the computed separation energies.
	In an infinite potential well, decreasing the width of the potential well will spread out the spectrum of single-particle energies, which depend on the radius as $1/R^2$, and we expect
	\begin{equation}\label{eq:dSdR}
	\delta S \approx -2\frac{\delta R}{R} S + {\rm const.} \,,
	\end{equation}
	where the radius shift $\delta R/R\approx -0.04$, and the constant shift depends on the potential well depth and any shifts thereof.
	We therefore anticipate that the residual $\delta S$ will exhibit a linear dependence on the separation energy $S$ with a slope of approximately $0.08$.
	Continuum and other effects will of course modify this simple picture.
	However, the angular momentum and Coulomb barriers, as well as the expansion on a harmonic oscillator basis keep deviations from this linear behavior within the error band, even down to slightly negative separation energies (see discussion below and Fig.~\ref{fig:Regression}).
	
	We therefore perform a Bayesian linear regression for the model
	\begin{equation}
	\delta S = \mathcal{A} S^{\rm th} + \mathcal{B} + \epsilon(\sigma^2)
	\end{equation}
	to obtain a posterior $p(\mathcal{A},\mathcal{B},\sigma^2 | S^{\rm th},S^{\rm exp})$ for the unknown parameters $\mathcal{A},\mathcal{B},\sigma^2$, given the theoretical and experimental data.
	We then marginalize over the posterior to obtain a posterior predictive distribution (PPD) $p(\tilde{S}^{\rm exp} |\tilde{S}^{\rm th}, S^{\rm th},S^{\rm exp})$ for a not-yet-measured separation energy $\tilde{S}^{\rm exp}$, given the theoretical value, and all the known data.
	Details can be found in the Supplemental Material.
	The result is shown in Fig.~\ref{fig:Regression}.
	As expected based on the discussion of Eq.~\eqref{eq:dSdR}, we obtain a slope of approximately 0.08 for each separation energy.
	Importantly, if we did not account for this systematic effect our uncertainty assessment would be biased by the relative abundance of data on well-bound isotopes.
	
	The cases requiring inconsistent valence spaces (which were not used in the regression) are marked with crosses in Fig.~\ref{fig:Regression}.
	In these cases the error due to incomplete cancellation of induced many-body effects is more difficult to model and so we are more conservative.
	For each valence-space boundary at $N,Z=8,20,40$ we apply an additional $S$-independent shift to the PPD---equal to the mean deviation from the regression line---and inflate the standard deviation by the size of the shift.
	
\begin{figure}
		\includegraphics[width=0.9\columnwidth]{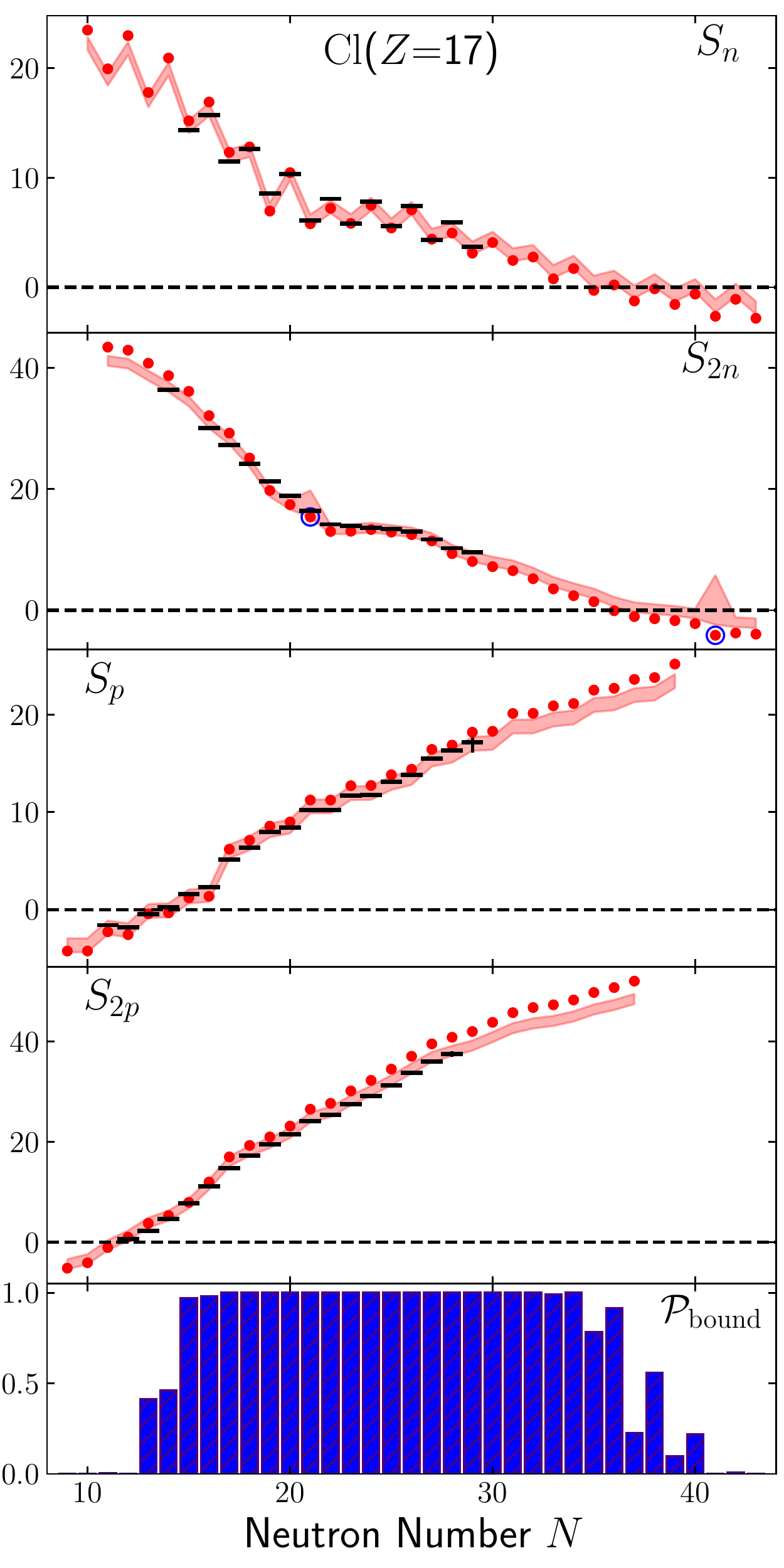}
		\caption{Separation energies and probability to be bound for the chlorine isotopes. The red dots indicate the results of the many-body calculation, while the red bands indicate the corrected 68\% uncertainty intervals. Blue circles indicate separation energies computed with inconsistent valence spaces.}
		\label{fig:cl_sep}
	\end{figure}

	The probability $\mathcal{P}_{1n}$ that an isotope is bound with respect to one-neutron emission is given by the fraction of the PPD for which $S_{n}$ is positive.
	The total probability to be bound is given by the fraction of the joint PPD for which all four separation energies are positive,
	\begin{equation}\label{eq:pbound}
	\mathcal{P}_{\rm bound} = \prod_{\alpha} \int_0^{\infty} d\tilde{S}_{\alpha}^{\rm exp} p(\tilde{S}_{\alpha}^{\rm exp} | \tilde{S}^{\rm th},S^{\rm th},S^{\rm exp}) \,,
	\end{equation}
	where as above $\alpha\in\{n,p,2n,2p\}$.
	As an illustration, the calculated separation energies, with the 68\% uncertainty band, are shown in Fig.~\ref{fig:cl_sep}, for chlorine isotopes.
	Analogous figures for all isotopic chains studied are included in the Supplemental Material, and a complete data table is provided as a Supplemental File.
	
	We translate this large-scale analysis into the main result of this Letter in Fig.~\ref{drip_unc}. 
	For each calculated nuclide from helium to iron, we assign a probability that it is bound with respect to one- or two-nucleon emission. 
	Every nuclide calculated is represented by a box in the plot, where its height and color denotes this probability: a full box is bound with probability 1, and an empty gray box is bound with probability 0.
	For ease of interpretation, we employ a color code with divisions at $\mathcal{P}_{\rm bound}=(0.05, 0.32, 0.68, 0.95)$.
	We denote experimentally known drip lines with a filled symbol and the heaviest (lightest) observed isotopes in the neutron-rich (deficient) regions with an open symbol. 
	
	Qualitatively, the location of the known drip lines appear to be reproduced well in Fig.~\ref{drip_unc}, both on the proton-rich side, and on the neutron-rich side where it is known up to $Z=10$.
	(Further quantitative validation of the approach is presented in Supplemental Material).
	Even the well known halo systems $^{11}$Li and $^{22}$C are predicted to be either bound or marginal, implying that physics of threshold systems lie within our estimated error bands.
	For all isotopic chains from sodium to chromium, our calculations indicate the likely existence of at least one isotope beyond the current known limits. 
	
	In calcium, earlier ab initio calculations have generally found that $^{62}$Ca~\cite{Hagen2013,Holt2014a} is the heaviest bound isotope (see also Ref.~\cite{Hergert2020}), with a very flat trend in binding energies beyond, leaving the location of the drip line ambiguous.
	The present analysis reflects that ambiguity; similar to oxygen the final bound nucleus could be closer to stability, but there is a reasonable probability that the drip line extends beyond $^{70}$Ca, as predicted in the statistical analysis of Ref.~\cite{Neuf2019}. 
	We note the remarkable similarities of the latter results to our ab initio predictions, which thus provides a consistent picture of the neutron drip line up to calcium from independent theoretical approaches.
	
	In summary, we have calculated ground-state energies of essentially all nuclei from helium to iron in the ab initio VS-IMSRG starting from NN and 3N interactions fit to few-body systems only. 
	Using available experimental data to quantify our theoretical error, we provide drip line predictions in the neutron-rich region above neon to guide ongoing and future efforts at rare-isotope beam facilities worldwide.
	This work also advances ab initio theory to global calculations, highlighting the rapidly increasing scope of the field, and the potential to provide predictions beyond where data exists with uncertainty estimates. 
	In principle, we might have expected a global survey to uncover deficiencies in the 1.8/2.0 (EM) interaction which were not apparent based on prior calculations of closed-shell nuclei or selected isotopic chains; however, we globally find an impressive agreement.
	While significant challenges remain in improving the rigor of theoretical error estimates from nuclear forces and many-body methods, the approach presented here indicates a path -- once current computational limitations can be overcome -- for ab initio input for nucleosynthesis calculations probing the r-process region of extreme neutron-rich nuclei. 
	
	\begin{acknowledgments}
		We would like to thank K.\,Fossez, H.\,Hergert, T.\,Miyagi, W.\,Nazarewicz, and J.\,K.\,Smith for enlightening discussions. TRIUMF receives funding via a contribution through the National Research Council Canada. This work was supported by NSERC, the U.S.~DOE under Contract DE-FG02-97ER41014, the Deutsche Forschungsgemeinschaft (DFG, German Research Foundation) -- Projektnummer 279384907 -- SFB 1245, the Cluster of Excellence ``Precision Physics, Fundamental Interactions, and Structure of Matter'' (PRISMA$^+$ EXC 2118/1) funded within the German Excellence Strategy (Project ID 39083149), and the BMBF under Contract No.~05P18RDFN1. Computations were performed with an allocation of computing resources at the J\"ulich Supercomputing Center and on the Oak Cluster at TRIUMF managed by the University of British Columbia department of Advanced Research Computing (ARC).
	\end{acknowledgments}
	
	\bibliographystyle{apsrev4-1}
	\bibliography{strongint,library}

\begin{thebibliography}{63}%
\makeatletter
\providecommand \@ifxundefined [1]{%
 \@ifx{#1\undefined}
}%
\providecommand \@ifnum [1]{%
 \ifnum #1\expandafter \@firstoftwo
 \else \expandafter \@secondoftwo
 \fi
}%
\providecommand \@ifx [1]{%
 \ifx #1\expandafter \@firstoftwo
 \else \expandafter \@secondoftwo
 \fi
}%
\providecommand \natexlab [1]{#1}%
\providecommand \enquote  [1]{``#1''}%
\providecommand \bibnamefont  [1]{#1}%
\providecommand \bibfnamefont [1]{#1}%
\providecommand \citenamefont [1]{#1}%
\providecommand \href@noop [0]{\@secondoftwo}%
\providecommand \href [0]{\begingroup \@sanitize@url \@href}%
\providecommand \@href[1]{\@@startlink{#1}\@@href}%
\providecommand \@@href[1]{\endgroup#1\@@endlink}%
\providecommand \@sanitize@url [0]{\catcode `\\12\catcode `\$12\catcode
  `\&12\catcode `\#12\catcode `\^12\catcode `\_12\catcode `\%12\relax}%
\providecommand \@@startlink[1]{}%
\providecommand \@@endlink[0]{}%
\providecommand \url  [0]{\begingroup\@sanitize@url \@url }%
\providecommand \@url [1]{\endgroup\@href {#1}{\urlprefix }}%
\providecommand \urlprefix  [0]{URL }%
\providecommand \Eprint [0]{\href }%
\providecommand \doibase [0]{http://dx.doi.org/}%
\providecommand \selectlanguage [0]{\@gobble}%
\providecommand \bibinfo  [0]{\@secondoftwo}%
\providecommand \bibfield  [0]{\@secondoftwo}%
\providecommand \translation [1]{[#1]}%
\providecommand \BibitemOpen [0]{}%
\providecommand \bibitemStop [0]{}%
\providecommand \bibitemNoStop [0]{.\EOS\space}%
\providecommand \EOS [0]{\spacefactor3000\relax}%
\providecommand \BibitemShut  [1]{\csname bibitem#1\endcsname}%
\let\auto@bib@innerbib\@empty
\bibitem [{\citenamefont {Thoennessen}(2016)}]{Thoe16Book}%
  \BibitemOpen
  \bibfield  {author} {\bibinfo {author} {\bibfnamefont {M.}~\bibnamefont
  {Thoennessen}},\ }\href {\doibase 10.1007/978-3-319-31763-2} {\emph {\bibinfo
  {title} {{The Discovery of Isotopes}}}}\ (\bibinfo  {publisher} {Springer},\
  \bibinfo {address} {New York},\ \bibinfo {year} {2016})\BibitemShut {NoStop}%
\bibitem [{\citenamefont {Ahn}\ \emph {et~al.}(2019)\citenamefont {Ahn},
  \citenamefont {Fukuda}, \citenamefont {Geissel}, \citenamefont {Inabe},
  \citenamefont {Iwasa}, \citenamefont {Kubo}, \citenamefont {Kusaka},
  \citenamefont {Morrissey}, \citenamefont {Murai}, \citenamefont {Nakamura},
  \citenamefont {Ohtake}, \citenamefont {Otsu}, \citenamefont {Sato},
  \citenamefont {Sherrill}, \citenamefont {Shimizu}, \citenamefont {Suzuki},
  \citenamefont {Takeda}, \citenamefont {Tarasov}, \citenamefont {Ueno},
  \citenamefont {Yanagisawa},\ and\ \citenamefont {Yoshida}}]{Ahn2019}%
  \BibitemOpen
  \bibfield  {author} {\bibinfo {author} {\bibfnamefont {D.~S.}\ \bibnamefont
  {Ahn}}, \bibinfo {author} {\bibfnamefont {N.}~\bibnamefont {Fukuda}},
  \bibinfo {author} {\bibfnamefont {H.}~\bibnamefont {Geissel}}, \bibinfo
  {author} {\bibfnamefont {N.}~\bibnamefont {Inabe}}, \bibinfo {author}
  {\bibfnamefont {N.}~\bibnamefont {Iwasa}}, \bibinfo {author} {\bibfnamefont
  {T.}~\bibnamefont {Kubo}}, \bibinfo {author} {\bibfnamefont {K.}~\bibnamefont
  {Kusaka}}, \bibinfo {author} {\bibfnamefont {D.~J.}\ \bibnamefont
  {Morrissey}}, \bibinfo {author} {\bibfnamefont {D.}~\bibnamefont {Murai}},
  \bibinfo {author} {\bibfnamefont {T.}~\bibnamefont {Nakamura}}, \bibinfo
  {author} {\bibfnamefont {M.}~\bibnamefont {Ohtake}}, \bibinfo {author}
  {\bibfnamefont {H.}~\bibnamefont {Otsu}}, \bibinfo {author} {\bibfnamefont
  {H.}~\bibnamefont {Sato}}, \bibinfo {author} {\bibfnamefont {B.~M.}\
  \bibnamefont {Sherrill}}, \bibinfo {author} {\bibfnamefont {Y.}~\bibnamefont
  {Shimizu}}, \bibinfo {author} {\bibfnamefont {H.}~\bibnamefont {Suzuki}},
  \bibinfo {author} {\bibfnamefont {H.}~\bibnamefont {Takeda}}, \bibinfo
  {author} {\bibfnamefont {O.~B.}\ \bibnamefont {Tarasov}}, \bibinfo {author}
  {\bibfnamefont {H.}~\bibnamefont {Ueno}}, \bibinfo {author} {\bibfnamefont
  {Y.}~\bibnamefont {Yanagisawa}}, \ and\ \bibinfo {author} {\bibfnamefont
  {K.}~\bibnamefont {Yoshida}},\ }\href {\doibase
  10.1103/PhysRevLett.123.212501} {\bibfield  {journal} {\bibinfo  {journal}
  {Phys. Rev. Lett.}\ }\textbf {\bibinfo {volume} {123}},\ \bibinfo {pages}
  {212501} (\bibinfo {year} {2019})}\BibitemShut {NoStop}%
\bibitem [{\citenamefont {Gade}\ and\ \citenamefont
  {Sherrill}(2016)}]{Gade16FRIB}%
  \BibitemOpen
  \bibfield  {author} {\bibinfo {author} {\bibfnamefont {A.}~\bibnamefont
  {Gade}}\ and\ \bibinfo {author} {\bibfnamefont {B.~M.}\ \bibnamefont
  {Sherrill}},\ }\href {\doibase 10.1088/0031-8949/91/5/053003} {\bibfield
  {journal} {\bibinfo  {journal} {Phys. Scripta}\ }\textbf {\bibinfo {volume}
  {91}},\ \bibinfo {pages} {053003} (\bibinfo {year} {2016})}\BibitemShut
  {NoStop}%
\bibitem [{\citenamefont {Durante}\ \emph {et~al.}(2019)\citenamefont
  {Durante}, \citenamefont {Indelicato}, \citenamefont {Jonson}, \citenamefont
  {Koch}, \citenamefont {Langanke}, \citenamefont {Mei{\ss}ner}, \citenamefont
  {Nappi}, \citenamefont {Nilsson}, \citenamefont {St{\"o}hlker},\ and\
  \citenamefont {Widmann}}]{Dura19FAIR}%
  \BibitemOpen
  \bibfield  {author} {\bibinfo {author} {\bibfnamefont {M.}~\bibnamefont
  {Durante}}, \bibinfo {author} {\bibfnamefont {P.}~\bibnamefont {Indelicato}},
  \bibinfo {author} {\bibfnamefont {B.}~\bibnamefont {Jonson}}, \bibinfo
  {author} {\bibfnamefont {V.}~\bibnamefont {Koch}}, \bibinfo {author}
  {\bibfnamefont {K.}~\bibnamefont {Langanke}}, \bibinfo {author}
  {\bibfnamefont {U.-G.}\ \bibnamefont {Mei{\ss}ner}}, \bibinfo {author}
  {\bibfnamefont {E.}~\bibnamefont {Nappi}}, \bibinfo {author} {\bibfnamefont
  {T.}~\bibnamefont {Nilsson}}, \bibinfo {author} {\bibfnamefont
  {T.}~\bibnamefont {St{\"o}hlker}}, \ and\ \bibinfo {author} {\bibfnamefont
  {E.}~\bibnamefont {Widmann}},\ }\href {\doibase 10.1088/1402-4896/aaf93f}
  {\bibfield  {journal} {\bibinfo  {journal} {Phys. Scripta}\ }\textbf
  {\bibinfo {volume} {94}},\ \bibinfo {pages} {033001} (\bibinfo {year}
  {2019})}\BibitemShut {NoStop}%
\bibitem [{\citenamefont {Tarasov}\ \emph {et~al.}(2018)\citenamefont
  {Tarasov}, \citenamefont {Ahn}, \citenamefont {Bazin}, \citenamefont
  {Fukuda}, \citenamefont {Gade}, \citenamefont {Hausmann}, \citenamefont
  {Inabe}, \citenamefont {Ishikawa}, \citenamefont {Iwasa}, \citenamefont
  {Kawata}, \citenamefont {Komatsubara}, \citenamefont {Kubo}, \citenamefont
  {Kusaka}, \citenamefont {Morrissey}, \citenamefont {Ohtake}, \citenamefont
  {Otsu}, \citenamefont {Portillo}, \citenamefont {Sakakibara}, \citenamefont
  {Sakurai}, \citenamefont {Sato}, \citenamefont {Sherrill}, \citenamefont
  {Shimizu}, \citenamefont {Stolz}, \citenamefont {Sumikama}, \citenamefont
  {Suzuki}, \citenamefont {Takeda}, \citenamefont {Thoennessen}, \citenamefont
  {Ueno}, \citenamefont {Yanagisawa},\ and\ \citenamefont
  {Yoshida}}]{Tara18Ca60}%
  \BibitemOpen
  \bibfield  {author} {\bibinfo {author} {\bibfnamefont {O.~B.}\ \bibnamefont
  {Tarasov}}, \bibinfo {author} {\bibfnamefont {D.~S.}\ \bibnamefont {Ahn}},
  \bibinfo {author} {\bibfnamefont {D.}~\bibnamefont {Bazin}}, \bibinfo
  {author} {\bibfnamefont {N.}~\bibnamefont {Fukuda}}, \bibinfo {author}
  {\bibfnamefont {A.}~\bibnamefont {Gade}}, \bibinfo {author} {\bibfnamefont
  {M.}~\bibnamefont {Hausmann}}, \bibinfo {author} {\bibfnamefont
  {N.}~\bibnamefont {Inabe}}, \bibinfo {author} {\bibfnamefont
  {S.}~\bibnamefont {Ishikawa}}, \bibinfo {author} {\bibfnamefont
  {N.}~\bibnamefont {Iwasa}}, \bibinfo {author} {\bibfnamefont
  {K.}~\bibnamefont {Kawata}}, \bibinfo {author} {\bibfnamefont
  {T.}~\bibnamefont {Komatsubara}}, \bibinfo {author} {\bibfnamefont
  {T.}~\bibnamefont {Kubo}}, \bibinfo {author} {\bibfnamefont {K.}~\bibnamefont
  {Kusaka}}, \bibinfo {author} {\bibfnamefont {D.~J.}\ \bibnamefont
  {Morrissey}}, \bibinfo {author} {\bibfnamefont {M.}~\bibnamefont {Ohtake}},
  \bibinfo {author} {\bibfnamefont {H.}~\bibnamefont {Otsu}}, \bibinfo {author}
  {\bibfnamefont {M.}~\bibnamefont {Portillo}}, \bibinfo {author}
  {\bibfnamefont {T.}~\bibnamefont {Sakakibara}}, \bibinfo {author}
  {\bibfnamefont {H.}~\bibnamefont {Sakurai}}, \bibinfo {author} {\bibfnamefont
  {H.}~\bibnamefont {Sato}}, \bibinfo {author} {\bibfnamefont {B.~M.}\
  \bibnamefont {Sherrill}}, \bibinfo {author} {\bibfnamefont {Y.}~\bibnamefont
  {Shimizu}}, \bibinfo {author} {\bibfnamefont {A.}~\bibnamefont {Stolz}},
  \bibinfo {author} {\bibfnamefont {T.}~\bibnamefont {Sumikama}}, \bibinfo
  {author} {\bibfnamefont {H.}~\bibnamefont {Suzuki}}, \bibinfo {author}
  {\bibfnamefont {H.}~\bibnamefont {Takeda}}, \bibinfo {author} {\bibfnamefont
  {M.}~\bibnamefont {Thoennessen}}, \bibinfo {author} {\bibfnamefont
  {H.}~\bibnamefont {Ueno}}, \bibinfo {author} {\bibfnamefont {Y.}~\bibnamefont
  {Yanagisawa}}, \ and\ \bibinfo {author} {\bibfnamefont {K.}~\bibnamefont
  {Yoshida}},\ }\href {\doibase 10.1103/PhysRevLett.121.022501} {\bibfield
  {journal} {\bibinfo  {journal} {Phys. Rev. Lett.}\ }\textbf {\bibinfo
  {volume} {121}},\ \bibinfo {pages} {022501} (\bibinfo {year}
  {2018})}\BibitemShut {NoStop}%
\bibitem [{\citenamefont {Mumpower}\ \emph {et~al.}(2016)\citenamefont
  {Mumpower}, \citenamefont {Surman}, \citenamefont {McLaughlin},\ and\
  \citenamefont {Aprahamian}}]{Mump15PPNP}%
  \BibitemOpen
  \bibfield  {author} {\bibinfo {author} {\bibfnamefont {M.~R.}\ \bibnamefont
  {Mumpower}}, \bibinfo {author} {\bibfnamefont {R.}~\bibnamefont {Surman}},
  \bibinfo {author} {\bibfnamefont {G.~C.}\ \bibnamefont {McLaughlin}}, \ and\
  \bibinfo {author} {\bibfnamefont {A.}~\bibnamefont {Aprahamian}},\ }\href
  {\doibase 10.1016/j.ppnp.2015.09.001} {\bibfield  {journal} {\bibinfo
  {journal} {Prog. Part. Nucl. Phys.}\ }\textbf {\bibinfo {volume} {86}},\
  \bibinfo {pages} {86} (\bibinfo {year} {2016})}\BibitemShut {NoStop}%
\bibitem [{\citenamefont {Martin}\ \emph {et~al.}(2016)\citenamefont {Martin},
  \citenamefont {Arcones}, \citenamefont {Nazarewicz},\ and\ \citenamefont
  {Olsen}}]{Mart16massrpro}%
  \BibitemOpen
  \bibfield  {author} {\bibinfo {author} {\bibfnamefont {D.}~\bibnamefont
  {Martin}}, \bibinfo {author} {\bibfnamefont {A.}~\bibnamefont {Arcones}},
  \bibinfo {author} {\bibfnamefont {W.}~\bibnamefont {Nazarewicz}}, \ and\
  \bibinfo {author} {\bibfnamefont {E.}~\bibnamefont {Olsen}},\ }\href
  {\doibase 10.1103/PhysRevLett.116.121101} {\bibfield  {journal} {\bibinfo
  {journal} {Phys. Rev. Lett.}\ }\textbf {\bibinfo {volume} {116}},\ \bibinfo
  {pages} {121101} (\bibinfo {year} {2016})}\BibitemShut {NoStop}%
\bibitem [{\citenamefont {Abbott}\ \emph {et~al.}(2017)\citenamefont {Abbott}
  \emph {et~al.}}]{Abbo17mm}%
  \BibitemOpen
  \bibfield  {author} {\bibinfo {author} {\bibfnamefont {B.~P.}\ \bibnamefont
  {Abbott}} \emph {et~al.},\ }\href {\doibase 10.3847/2041-8213/aa91c9}
  {\bibfield  {journal} {\bibinfo  {journal} {Astrophys. J.}\ }\textbf
  {\bibinfo {volume} {848}},\ \bibinfo {pages} {L12} (\bibinfo {year}
  {2017})}\BibitemShut {NoStop}%
\bibitem [{\citenamefont {Erler}\ \emph {et~al.}(2012)\citenamefont {Erler},
  \citenamefont {Birge}, \citenamefont {Kortelainen}, \citenamefont
  {Nazarewicz}, \citenamefont {Olsen}, \citenamefont {Perhac},\ and\
  \citenamefont {Stoitsov}}]{Erle12Nature}%
  \BibitemOpen
  \bibfield  {author} {\bibinfo {author} {\bibfnamefont {J.}~\bibnamefont
  {Erler}}, \bibinfo {author} {\bibfnamefont {N.}~\bibnamefont {Birge}},
  \bibinfo {author} {\bibfnamefont {M.}~\bibnamefont {Kortelainen}}, \bibinfo
  {author} {\bibfnamefont {W.}~\bibnamefont {Nazarewicz}}, \bibinfo {author}
  {\bibfnamefont {E.}~\bibnamefont {Olsen}}, \bibinfo {author} {\bibfnamefont
  {A.~M.}\ \bibnamefont {Perhac}}, \ and\ \bibinfo {author} {\bibfnamefont
  {M.}~\bibnamefont {Stoitsov}},\ }\href {\doibase 10.1038/nature11188}
  {\bibfield  {journal} {\bibinfo  {journal} {Nature}\ }\textbf {\bibinfo
  {volume} {486}},\ \bibinfo {pages} {509} (\bibinfo {year}
  {2012})}\BibitemShut {NoStop}%
\bibitem [{\citenamefont {Neufcourt}\ \emph {et~al.}(2019)\citenamefont
  {Neufcourt}, \citenamefont {Cao}, \citenamefont {Nazarewicz}, \citenamefont
  {Olsen},\ and\ \citenamefont {Viens}}]{Neuf2019}%
  \BibitemOpen
  \bibfield  {author} {\bibinfo {author} {\bibfnamefont {L.}~\bibnamefont
  {Neufcourt}}, \bibinfo {author} {\bibfnamefont {Y.}~\bibnamefont {Cao}},
  \bibinfo {author} {\bibfnamefont {W.}~\bibnamefont {Nazarewicz}}, \bibinfo
  {author} {\bibfnamefont {E.}~\bibnamefont {Olsen}}, \ and\ \bibinfo {author}
  {\bibfnamefont {F.}~\bibnamefont {Viens}},\ }\href {\doibase
  10.1103/PhysRevLett.122.062502} {\bibfield  {journal} {\bibinfo  {journal}
  {Phys. Rev. Lett.}\ }\textbf {\bibinfo {volume} {122}},\ \bibinfo {pages}
  {062502} (\bibinfo {year} {2019})}\BibitemShut {NoStop}%
\bibitem [{\citenamefont {Yoshida}\ \emph {et~al.}(2018)\citenamefont
  {Yoshida}, \citenamefont {Shimizu}, \citenamefont {Togashi},\ and\
  \citenamefont {Otsuka}}]{Yoshida18}%
  \BibitemOpen
  \bibfield  {author} {\bibinfo {author} {\bibfnamefont {S.}~\bibnamefont
  {Yoshida}}, \bibinfo {author} {\bibfnamefont {N.}~\bibnamefont {Shimizu}},
  \bibinfo {author} {\bibfnamefont {T.}~\bibnamefont {Togashi}}, \ and\
  \bibinfo {author} {\bibfnamefont {T.}~\bibnamefont {Otsuka}},\ }\href
  {\doibase 10.1103/PhysRevC.98.061301} {\bibfield  {journal} {\bibinfo
  {journal} {Phys. Rev. C}\ }\textbf {\bibinfo {volume} {98}},\ \bibinfo
  {pages} {061301(R)} (\bibinfo {year} {2018})}\BibitemShut {NoStop}%
\bibitem [{\citenamefont {Epelbaum}\ \emph {et~al.}(2009)\citenamefont
  {Epelbaum}, \citenamefont {Hammer},\ and\ \citenamefont
  {Mei{\ss}ner}}]{Epel09RMP}%
  \BibitemOpen
  \bibfield  {author} {\bibinfo {author} {\bibfnamefont {E.}~\bibnamefont
  {Epelbaum}}, \bibinfo {author} {\bibfnamefont {H.-W.}\ \bibnamefont
  {Hammer}}, \ and\ \bibinfo {author} {\bibfnamefont {U.-G.}\ \bibnamefont
  {Mei{\ss}ner}},\ }\href {\doibase 10.1103/RevModPhys.81.1773} {\bibfield
  {journal} {\bibinfo  {journal} {Rev. Mod. Phys.}\ }\textbf {\bibinfo {volume}
  {81}},\ \bibinfo {pages} {1773} (\bibinfo {year} {2009})}\BibitemShut
  {NoStop}%
\bibitem [{\citenamefont {Machleidt}\ and\ \citenamefont
  {Entem}(2011)}]{Mach11PR}%
  \BibitemOpen
  \bibfield  {author} {\bibinfo {author} {\bibfnamefont {R.}~\bibnamefont
  {Machleidt}}\ and\ \bibinfo {author} {\bibfnamefont {D.~R.}\ \bibnamefont
  {Entem}},\ }\href {\doibase 10.1016/j.physrep.2011.02.001} {\bibfield
  {journal} {\bibinfo  {journal} {Phys. Rep.}\ }\textbf {\bibinfo {volume}
  {503}},\ \bibinfo {pages} {1} (\bibinfo {year} {2011})}\BibitemShut {NoStop}%
\bibitem [{\citenamefont {Hammer}\ \emph {et~al.}(2013)\citenamefont {Hammer},
  \citenamefont {Nogga},\ and\ \citenamefont {Schwenk}}]{Hamm13RMP}%
  \BibitemOpen
  \bibfield  {author} {\bibinfo {author} {\bibfnamefont {H.-W.}\ \bibnamefont
  {Hammer}}, \bibinfo {author} {\bibfnamefont {A.}~\bibnamefont {Nogga}}, \
  and\ \bibinfo {author} {\bibfnamefont {A.}~\bibnamefont {Schwenk}},\ }\href
  {\doibase 10.1103/RevModPhys.85.197} {\bibfield  {journal} {\bibinfo
  {journal} {Rev. Mod. Phys.}\ }\textbf {\bibinfo {volume} {85}},\ \bibinfo
  {pages} {197} (\bibinfo {year} {2013})}\BibitemShut {NoStop}%
\bibitem [{\citenamefont {Bogner}\ \emph {et~al.}(2007)\citenamefont {Bogner},
  \citenamefont {Furnstahl},\ and\ \citenamefont {Perry}}]{Bogn07SRG}%
  \BibitemOpen
  \bibfield  {author} {\bibinfo {author} {\bibfnamefont {S.~K.}\ \bibnamefont
  {Bogner}}, \bibinfo {author} {\bibfnamefont {R.~J.}\ \bibnamefont
  {Furnstahl}}, \ and\ \bibinfo {author} {\bibfnamefont {R.~J.}\ \bibnamefont
  {Perry}},\ }\href {\doibase 10.1103/PhysRevC.75.061001} {\bibfield  {journal}
  {\bibinfo  {journal} {Phys. Rev. C}\ }\textbf {\bibinfo {volume} {75}},\
  \bibinfo {pages} {061001(R)} (\bibinfo {year} {2007})}\BibitemShut {NoStop}%
\bibitem [{\citenamefont {Bogner}\ \emph {et~al.}(2010)\citenamefont {Bogner},
  \citenamefont {Furnstahl},\ and\ \citenamefont {Schwenk}}]{Bogn10PPNP}%
  \BibitemOpen
  \bibfield  {author} {\bibinfo {author} {\bibfnamefont {S.~K.}\ \bibnamefont
  {Bogner}}, \bibinfo {author} {\bibfnamefont {R.~J.}\ \bibnamefont
  {Furnstahl}}, \ and\ \bibinfo {author} {\bibfnamefont {A.}~\bibnamefont
  {Schwenk}},\ }\href {\doibase 10.1016/j.ppnp.2010.03.001} {\bibfield
  {journal} {\bibinfo  {journal} {Prog. Part. Nucl. Phys.}\ }\textbf {\bibinfo
  {volume} {65}},\ \bibinfo {pages} {94} (\bibinfo {year} {2010})}\BibitemShut
  {NoStop}%
\bibitem [{\citenamefont {Hebeler}\ \emph {et~al.}(2011)\citenamefont
  {Hebeler}, \citenamefont {Bogner}, \citenamefont {Furnstahl}, \citenamefont
  {Nogga},\ and\ \citenamefont {Schwenk}}]{Hebe11fits}%
  \BibitemOpen
  \bibfield  {author} {\bibinfo {author} {\bibfnamefont {K.}~\bibnamefont
  {Hebeler}}, \bibinfo {author} {\bibfnamefont {S.~K.}\ \bibnamefont {Bogner}},
  \bibinfo {author} {\bibfnamefont {R.~J.}\ \bibnamefont {Furnstahl}}, \bibinfo
  {author} {\bibfnamefont {A.}~\bibnamefont {Nogga}}, \ and\ \bibinfo {author}
  {\bibfnamefont {A.}~\bibnamefont {Schwenk}},\ }\href {\doibase
  10.1103/PhysRevC.83.031301} {\bibfield  {journal} {\bibinfo  {journal} {Phys.
  Rev. C}\ }\textbf {\bibinfo {volume} {83}},\ \bibinfo {pages} {031301(R)}
  (\bibinfo {year} {2011})}\BibitemShut {NoStop}%
\bibitem [{\citenamefont {Ekstr\"om}\ \emph {et~al.}(2015)\citenamefont
  {Ekstr\"om}, \citenamefont {Jansen}, \citenamefont {Wendt}, \citenamefont
  {Hagen}, \citenamefont {Papenbrock}, \citenamefont {Carlsson}, \citenamefont
  {Forss\'en}, \citenamefont {Hjorth-Jensen}, \citenamefont {Navr\'atil},\ and\
  \citenamefont {Nazarewicz}}]{Ekst15sat}%
  \BibitemOpen
  \bibfield  {author} {\bibinfo {author} {\bibfnamefont {A.}~\bibnamefont
  {Ekstr\"om}}, \bibinfo {author} {\bibfnamefont {G.~R.}\ \bibnamefont
  {Jansen}}, \bibinfo {author} {\bibfnamefont {K.~A.}\ \bibnamefont {Wendt}},
  \bibinfo {author} {\bibfnamefont {G.}~\bibnamefont {Hagen}}, \bibinfo
  {author} {\bibfnamefont {T.}~\bibnamefont {Papenbrock}}, \bibinfo {author}
  {\bibfnamefont {B.~D.}\ \bibnamefont {Carlsson}}, \bibinfo {author}
  {\bibfnamefont {C.}~\bibnamefont {Forss\'en}}, \bibinfo {author}
  {\bibfnamefont {M.}~\bibnamefont {Hjorth-Jensen}}, \bibinfo {author}
  {\bibfnamefont {P.}~\bibnamefont {Navr\'atil}}, \ and\ \bibinfo {author}
  {\bibfnamefont {W.}~\bibnamefont {Nazarewicz}},\ }\href {\doibase
  10.1103/PhysRevC.91.051301} {\bibfield  {journal} {\bibinfo  {journal} {Phys.
  Rev. C}\ }\textbf {\bibinfo {volume} {91}},\ \bibinfo {pages} {051301(R)}
  (\bibinfo {year} {2015})}\BibitemShut {NoStop}%
\bibitem [{\citenamefont {Morris}\ \emph {et~al.}(2018)\citenamefont {Morris},
  \citenamefont {Simonis}, \citenamefont {Stroberg}, \citenamefont {Stumpf},
  \citenamefont {Hagen}, \citenamefont {Holt}, \citenamefont {Jansen},
  \citenamefont {Papenbrock}, \citenamefont {Roth},\ and\ \citenamefont
  {Schwenk}}]{Morr17Tin}%
  \BibitemOpen
  \bibfield  {author} {\bibinfo {author} {\bibfnamefont {T.~D.}\ \bibnamefont
  {Morris}}, \bibinfo {author} {\bibfnamefont {J.}~\bibnamefont {Simonis}},
  \bibinfo {author} {\bibfnamefont {S.~R.}\ \bibnamefont {Stroberg}}, \bibinfo
  {author} {\bibfnamefont {C.}~\bibnamefont {Stumpf}}, \bibinfo {author}
  {\bibfnamefont {G.}~\bibnamefont {Hagen}}, \bibinfo {author} {\bibfnamefont
  {J.~D.}\ \bibnamefont {Holt}}, \bibinfo {author} {\bibfnamefont {G.~R.}\
  \bibnamefont {Jansen}}, \bibinfo {author} {\bibfnamefont {T.}~\bibnamefont
  {Papenbrock}}, \bibinfo {author} {\bibfnamefont {R.}~\bibnamefont {Roth}}, \
  and\ \bibinfo {author} {\bibfnamefont {A.}~\bibnamefont {Schwenk}},\ }\href
  {\doibase 10.1103/PhysRevLett.120.152503} {\bibfield  {journal} {\bibinfo
  {journal} {Phys. Rev. Lett.}\ }\textbf {\bibinfo {volume} {120}},\ \bibinfo
  {pages} {152503} (\bibinfo {year} {2018})}\BibitemShut {NoStop}%
\bibitem [{\citenamefont {Otsuka}\ \emph {et~al.}(2010)\citenamefont {Otsuka},
  \citenamefont {Suzuki}, \citenamefont {Holt}, \citenamefont {Schwenk},\ and\
  \citenamefont {Akaishi}}]{Otsu10Ox}%
  \BibitemOpen
  \bibfield  {author} {\bibinfo {author} {\bibfnamefont {T.}~\bibnamefont
  {Otsuka}}, \bibinfo {author} {\bibfnamefont {T.}~\bibnamefont {Suzuki}},
  \bibinfo {author} {\bibfnamefont {J.~D.}\ \bibnamefont {Holt}}, \bibinfo
  {author} {\bibfnamefont {A.}~\bibnamefont {Schwenk}}, \ and\ \bibinfo
  {author} {\bibfnamefont {Y.}~\bibnamefont {Akaishi}},\ }\href {\doibase
  10.1103/PhysRevLett.105.032501} {\bibfield  {journal} {\bibinfo  {journal}
  {Phys. Rev. Lett.}\ }\textbf {\bibinfo {volume} {105}},\ \bibinfo {pages}
  {032501} (\bibinfo {year} {2010})}\BibitemShut {NoStop}%
\bibitem [{\citenamefont {Hagen}\ \emph {et~al.}(2012)\citenamefont {Hagen},
  \citenamefont {Hjorth-Jensen}, \citenamefont {Jansen}, \citenamefont
  {Machleidt},\ and\ \citenamefont {Papenbrock}}]{Hage12Ox3N}%
  \BibitemOpen
  \bibfield  {author} {\bibinfo {author} {\bibfnamefont {G.}~\bibnamefont
  {Hagen}}, \bibinfo {author} {\bibfnamefont {M.}~\bibnamefont
  {Hjorth-Jensen}}, \bibinfo {author} {\bibfnamefont {G.~R.}\ \bibnamefont
  {Jansen}}, \bibinfo {author} {\bibfnamefont {R.}~\bibnamefont {Machleidt}}, \
  and\ \bibinfo {author} {\bibfnamefont {T.}~\bibnamefont {Papenbrock}},\
  }\href {\doibase 10.1103/PhysRevLett.108.242501} {\bibfield  {journal}
  {\bibinfo  {journal} {Phys. Rev. Lett.}\ }\textbf {\bibinfo {volume} {108}},\
  \bibinfo {pages} {242501} (\bibinfo {year} {2012})}\BibitemShut {NoStop}%
\bibitem [{\citenamefont {Cipollone}\ \emph {et~al.}(2013)\citenamefont
  {Cipollone}, \citenamefont {Barbieri},\ and\ \citenamefont
  {Navr{\'a}til}}]{Cipo13Ox}%
  \BibitemOpen
  \bibfield  {author} {\bibinfo {author} {\bibfnamefont {A.}~\bibnamefont
  {Cipollone}}, \bibinfo {author} {\bibfnamefont {C.}~\bibnamefont {Barbieri}},
  \ and\ \bibinfo {author} {\bibfnamefont {P.}~\bibnamefont {Navr{\'a}til}},\
  }\href {\doibase 10.1103/PhysRevLett.111.062501} {\bibfield  {journal}
  {\bibinfo  {journal} {Phys. Rev. Lett.}\ }\textbf {\bibinfo {volume} {111}},\
  \bibinfo {pages} {062501} (\bibinfo {year} {2013})}\BibitemShut {NoStop}%
\bibitem [{\citenamefont {Hergert}\ \emph {et~al.}(2013)\citenamefont
  {Hergert}, \citenamefont {Binder}, \citenamefont {Calci}, \citenamefont
  {Langhammer},\ and\ \citenamefont {Roth}}]{Herg13MR}%
  \BibitemOpen
  \bibfield  {author} {\bibinfo {author} {\bibfnamefont {H.}~\bibnamefont
  {Hergert}}, \bibinfo {author} {\bibfnamefont {S.}~\bibnamefont {Binder}},
  \bibinfo {author} {\bibfnamefont {A.}~\bibnamefont {Calci}}, \bibinfo
  {author} {\bibfnamefont {J.}~\bibnamefont {Langhammer}}, \ and\ \bibinfo
  {author} {\bibfnamefont {R.}~\bibnamefont {Roth}},\ }\href {\doibase
  10.1103/PhysRevLett.110.242501} {\bibfield  {journal} {\bibinfo  {journal}
  {Phys. Rev. Lett.}\ }\textbf {\bibinfo {volume} {110}},\ \bibinfo {pages}
  {242501} (\bibinfo {year} {2013})}\BibitemShut {NoStop}%
\bibitem [{\citenamefont {Holt}\ \emph {et~al.}(2013)\citenamefont {Holt},
  \citenamefont {Men{\'e}ndez},\ and\ \citenamefont {Schwenk}}]{Holt13PR}%
  \BibitemOpen
  \bibfield  {author} {\bibinfo {author} {\bibfnamefont {J.~D.}\ \bibnamefont
  {Holt}}, \bibinfo {author} {\bibfnamefont {J.}~\bibnamefont {Men{\'e}ndez}},
  \ and\ \bibinfo {author} {\bibfnamefont {A.}~\bibnamefont {Schwenk}},\ }\href
  {\doibase 10.1103/PhysRevLett.110.022502} {\bibfield  {journal} {\bibinfo
  {journal} {Phys. Rev. Lett.}\ }\textbf {\bibinfo {volume} {110}},\ \bibinfo
  {pages} {022502} (\bibinfo {year} {2013})}\BibitemShut {NoStop}%
\bibitem [{\citenamefont {Hebeler}\ \emph {et~al.}(2015)\citenamefont
  {Hebeler}, \citenamefont {Holt}, \citenamefont {Men{\'e}ndez},\ and\
  \citenamefont {Schwenk}}]{Hebe15ARNPS}%
  \BibitemOpen
  \bibfield  {author} {\bibinfo {author} {\bibfnamefont {K.}~\bibnamefont
  {Hebeler}}, \bibinfo {author} {\bibfnamefont {J.~D.}\ \bibnamefont {Holt}},
  \bibinfo {author} {\bibfnamefont {J.}~\bibnamefont {Men{\'e}ndez}}, \ and\
  \bibinfo {author} {\bibfnamefont {A.}~\bibnamefont {Schwenk}},\ }\href
  {\doibase 10.1146/annurev-nucl-102313-025446} {\bibfield  {journal} {\bibinfo
   {journal} {Annu. Rev. Nucl. Part. Sci.}\ }\textbf {\bibinfo {volume} {65}},\
  \bibinfo {pages} {457} (\bibinfo {year} {2015})}\BibitemShut {NoStop}%
\bibitem [{\citenamefont {Barrett}\ \emph {et~al.}(2013)\citenamefont
  {Barrett}, \citenamefont {Navr{\'a}til},\ and\ \citenamefont
  {Vary}}]{Barr13PPNP}%
  \BibitemOpen
  \bibfield  {author} {\bibinfo {author} {\bibfnamefont {B.~R.}\ \bibnamefont
  {Barrett}}, \bibinfo {author} {\bibfnamefont {P.}~\bibnamefont
  {Navr{\'a}til}}, \ and\ \bibinfo {author} {\bibfnamefont {J.~P.}\
  \bibnamefont {Vary}},\ }\href {\doibase 10.1016/j.ppnp.2012.10.003}
  {\bibfield  {journal} {\bibinfo  {journal} {Prog. Part. Nucl. Phys.}\
  }\textbf {\bibinfo {volume} {69}},\ \bibinfo {pages} {131} (\bibinfo {year}
  {2013})}\BibitemShut {NoStop}%
\bibitem [{\citenamefont {Hagen}\ \emph {et~al.}(2014)\citenamefont {Hagen},
  \citenamefont {Papenbrock}, \citenamefont {Hjorth-Jensen},\ and\
  \citenamefont {Dean}}]{Hage14RPP}%
  \BibitemOpen
  \bibfield  {author} {\bibinfo {author} {\bibfnamefont {G.}~\bibnamefont
  {Hagen}}, \bibinfo {author} {\bibfnamefont {T.}~\bibnamefont {Papenbrock}},
  \bibinfo {author} {\bibfnamefont {M.}~\bibnamefont {Hjorth-Jensen}}, \ and\
  \bibinfo {author} {\bibfnamefont {D.~J.}\ \bibnamefont {Dean}},\ }\href
  {\doibase 10.1088/0034-4885/77/9/096302} {\bibfield  {journal} {\bibinfo
  {journal} {Rep. Prog. Phys.}\ }\textbf {\bibinfo {volume} {77}},\ \bibinfo
  {pages} {096302} (\bibinfo {year} {2014})}\BibitemShut {NoStop}%
\bibitem [{\citenamefont {Carlson}\ \emph {et~al.}(2015)\citenamefont
  {Carlson}, \citenamefont {Gandolfi}, \citenamefont {Pederiva}, \citenamefont
  {Pieper}, \citenamefont {Schiavilla}, \citenamefont {Schmidt},\ and\
  \citenamefont {Wiringa}}]{Carl15RMP}%
  \BibitemOpen
  \bibfield  {author} {\bibinfo {author} {\bibfnamefont {J.}~\bibnamefont
  {Carlson}}, \bibinfo {author} {\bibfnamefont {S.}~\bibnamefont {Gandolfi}},
  \bibinfo {author} {\bibfnamefont {F.}~\bibnamefont {Pederiva}}, \bibinfo
  {author} {\bibfnamefont {S.~C.}\ \bibnamefont {Pieper}}, \bibinfo {author}
  {\bibfnamefont {R.}~\bibnamefont {Schiavilla}}, \bibinfo {author}
  {\bibfnamefont {K.~E.}\ \bibnamefont {Schmidt}}, \ and\ \bibinfo {author}
  {\bibfnamefont {R.~B.}\ \bibnamefont {Wiringa}},\ }\href {\doibase
  10.1103/RevModPhys.87.1067} {\bibfield  {journal} {\bibinfo  {journal} {Rev.
  Mod. Phys.}\ }\textbf {\bibinfo {volume} {87}},\ \bibinfo {pages} {1067}
  (\bibinfo {year} {2015})}\BibitemShut {NoStop}%
\bibitem [{\citenamefont {Hergert}\ \emph {et~al.}(2016)\citenamefont
  {Hergert}, \citenamefont {Bogner}, \citenamefont {Morris}, \citenamefont
  {Schwenk},\ and\ \citenamefont {Tsukiyama}}]{Herg16PR}%
  \BibitemOpen
  \bibfield  {author} {\bibinfo {author} {\bibfnamefont {H.}~\bibnamefont
  {Hergert}}, \bibinfo {author} {\bibfnamefont {S.~K.}\ \bibnamefont {Bogner}},
  \bibinfo {author} {\bibfnamefont {T.~D.}\ \bibnamefont {Morris}}, \bibinfo
  {author} {\bibfnamefont {A.}~\bibnamefont {Schwenk}}, \ and\ \bibinfo
  {author} {\bibfnamefont {K.}~\bibnamefont {Tsukiyama}},\ }\href {\doibase
  10.1016/j.physrep.2015.12.007} {\bibfield  {journal} {\bibinfo  {journal}
  {Phys. Rept.}\ }\textbf {\bibinfo {volume} {621}},\ \bibinfo {pages} {165}
  (\bibinfo {year} {2016})}\BibitemShut {NoStop}%
\bibitem [{\citenamefont {Stroberg}\ \emph {et~al.}(2019)\citenamefont
  {Stroberg}, \citenamefont {Bogner}, \citenamefont {Hergert},\ and\
  \citenamefont {Holt}}]{Stroberg19ARNPS}%
  \BibitemOpen
  \bibfield  {author} {\bibinfo {author} {\bibfnamefont {S.~R.}\ \bibnamefont
  {Stroberg}}, \bibinfo {author} {\bibfnamefont {S.~K.}\ \bibnamefont
  {Bogner}}, \bibinfo {author} {\bibfnamefont {H.}~\bibnamefont {Hergert}}, \
  and\ \bibinfo {author} {\bibfnamefont {J.~D.}\ \bibnamefont {Holt}},\ }\href
  {\doibase 10.1146/annurev-nucl-101917-021120} {\bibfield  {journal} {\bibinfo
   {journal} {Annu. Rev. Nucl. Part. Sci.}\ }\textbf {\bibinfo {volume} {69}},\
  \bibinfo {pages} {307} (\bibinfo {year} {2019})},\ \Eprint
  {http://arxiv.org/abs/1902.06154} {1902.06154} \BibitemShut {NoStop}%
\bibitem [{\citenamefont {Barbieri}\ and\ \citenamefont
  {Carbone}(2017)}]{Barb17SCGFlectnote}%
  \BibitemOpen
  \bibfield  {author} {\bibinfo {author} {\bibfnamefont {C.}~\bibnamefont
  {Barbieri}}\ and\ \bibinfo {author} {\bibfnamefont {A.}~\bibnamefont
  {Carbone}},\ }\enquote {\bibinfo {title} {Self-consistent green's function
  approaches},}\ in\ \href {\doibase 10.1007/978-3-319-53336-0_11} {\emph
  {\bibinfo {booktitle} {An Advanced Course in Computational Nuclear Physics:
  Bridging the Scales from Quarks to Neutron Stars}}},\ \bibinfo {editor}
  {edited by\ \bibinfo {editor} {\bibfnamefont {M.}~\bibnamefont
  {Hjorth-Jensen}}, \bibinfo {editor} {\bibfnamefont {M.~P.}\ \bibnamefont
  {Lombardo}}, \ and\ \bibinfo {editor} {\bibfnamefont {U.}~\bibnamefont {van
  Kolck}}}\ (\bibinfo  {publisher} {Springer International Publishing},\
  \bibinfo {address} {Cham},\ \bibinfo {year} {2017})\ pp.\ \bibinfo {pages}
  {571--644}\BibitemShut {NoStop}%
\bibitem [{\citenamefont {Bogner}\ \emph {et~al.}(2014)\citenamefont {Bogner},
  \citenamefont {Hergert}, \citenamefont {Holt}, \citenamefont {Schwenk},
  \citenamefont {Binder}, \citenamefont {Calci}, \citenamefont {Langhammer},\
  and\ \citenamefont {Roth}}]{Bogn14SM}%
  \BibitemOpen
  \bibfield  {author} {\bibinfo {author} {\bibfnamefont {S.~K.}\ \bibnamefont
  {Bogner}}, \bibinfo {author} {\bibfnamefont {H.}~\bibnamefont {Hergert}},
  \bibinfo {author} {\bibfnamefont {J.~D.}\ \bibnamefont {Holt}}, \bibinfo
  {author} {\bibfnamefont {A.}~\bibnamefont {Schwenk}}, \bibinfo {author}
  {\bibfnamefont {S.}~\bibnamefont {Binder}}, \bibinfo {author} {\bibfnamefont
  {A.}~\bibnamefont {Calci}}, \bibinfo {author} {\bibfnamefont
  {J.}~\bibnamefont {Langhammer}}, \ and\ \bibinfo {author} {\bibfnamefont
  {R.}~\bibnamefont {Roth}},\ }\href {\doibase 10.1103/PhysRevLett.113.142501}
  {\bibfield  {journal} {\bibinfo  {journal} {Phys. Rev. Lett.}\ }\textbf
  {\bibinfo {volume} {113}},\ \bibinfo {pages} {142501} (\bibinfo {year}
  {2014})}\BibitemShut {NoStop}%
\bibitem [{\citenamefont {Jansen}\ \emph {et~al.}(2014)\citenamefont {Jansen},
  \citenamefont {Engel}, \citenamefont {Hagen}, \citenamefont {Navr{\'a}til},\
  and\ \citenamefont {Signoracci}}]{Jans14SM}%
  \BibitemOpen
  \bibfield  {author} {\bibinfo {author} {\bibfnamefont {G.~R.}\ \bibnamefont
  {Jansen}}, \bibinfo {author} {\bibfnamefont {J.}~\bibnamefont {Engel}},
  \bibinfo {author} {\bibfnamefont {G.}~\bibnamefont {Hagen}}, \bibinfo
  {author} {\bibfnamefont {P.}~\bibnamefont {Navr{\'a}til}}, \ and\ \bibinfo
  {author} {\bibfnamefont {A.}~\bibnamefont {Signoracci}},\ }\href {\doibase
  10.1103/PhysRevLett.113.142502} {\bibfield  {journal} {\bibinfo  {journal}
  {Phys. Rev. Lett.}\ }\textbf {\bibinfo {volume} {113}},\ \bibinfo {pages}
  {142502} (\bibinfo {year} {2014})}\BibitemShut {NoStop}%
\bibitem [{\citenamefont {Som{\`a}}\ \emph {et~al.}(2014)\citenamefont
  {Som{\`a}}, \citenamefont {Cipollone}, \citenamefont {Barbieri},
  \citenamefont {Navr{\'a}til},\ and\ \citenamefont {Duguet}}]{Soma14GGF2N3N}%
  \BibitemOpen
  \bibfield  {author} {\bibinfo {author} {\bibfnamefont {V.}~\bibnamefont
  {Som{\`a}}}, \bibinfo {author} {\bibfnamefont {A.}~\bibnamefont {Cipollone}},
  \bibinfo {author} {\bibfnamefont {C.}~\bibnamefont {Barbieri}}, \bibinfo
  {author} {\bibfnamefont {P.}~\bibnamefont {Navr{\'a}til}}, \ and\ \bibinfo
  {author} {\bibfnamefont {T.}~\bibnamefont {Duguet}},\ }\href {\doibase
  10.1103/PhysRevC.89.061301} {\bibfield  {journal} {\bibinfo  {journal} {Phys.
  Rev. C}\ }\textbf {\bibinfo {volume} {89}},\ \bibinfo {pages} {061301(R)}
  (\bibinfo {year} {2014})}\BibitemShut {NoStop}%
\bibitem [{\citenamefont {Stroberg}\ \emph {et~al.}(2017)\citenamefont
  {Stroberg}, \citenamefont {Calci}, \citenamefont {Hergert}, \citenamefont
  {Holt}, \citenamefont {Bogner}, \citenamefont {Roth},\ and\ \citenamefont
  {Schwenk}}]{Stro17ENO}%
  \BibitemOpen
  \bibfield  {author} {\bibinfo {author} {\bibfnamefont {S.~R.}\ \bibnamefont
  {Stroberg}}, \bibinfo {author} {\bibfnamefont {A.}~\bibnamefont {Calci}},
  \bibinfo {author} {\bibfnamefont {H.}~\bibnamefont {Hergert}}, \bibinfo
  {author} {\bibfnamefont {J.~D.}\ \bibnamefont {Holt}}, \bibinfo {author}
  {\bibfnamefont {S.~K.}\ \bibnamefont {Bogner}}, \bibinfo {author}
  {\bibfnamefont {R.}~\bibnamefont {Roth}}, \ and\ \bibinfo {author}
  {\bibfnamefont {A.}~\bibnamefont {Schwenk}},\ }\href {\doibase
  10.1103/PhysRevLett.118.032502} {\bibfield  {journal} {\bibinfo  {journal}
  {Phys. Rev. Lett.}\ }\textbf {\bibinfo {volume} {118}},\ \bibinfo {pages}
  {032502} (\bibinfo {year} {2017})}\BibitemShut {NoStop}%
\bibitem [{\citenamefont {Tsukiyama}\ \emph {et~al.}(2012)\citenamefont
  {Tsukiyama}, \citenamefont {Bogner},\ and\ \citenamefont
  {Schwenk}}]{Tsuk12SM}%
  \BibitemOpen
  \bibfield  {author} {\bibinfo {author} {\bibfnamefont {K.}~\bibnamefont
  {Tsukiyama}}, \bibinfo {author} {\bibfnamefont {S.~K.}\ \bibnamefont
  {Bogner}}, \ and\ \bibinfo {author} {\bibfnamefont {A.}~\bibnamefont
  {Schwenk}},\ }\href {\doibase 10.1103/PhysRevC.85.061304} {\bibfield
  {journal} {\bibinfo  {journal} {Phys. Rev. C}\ }\textbf {\bibinfo {volume}
  {85}},\ \bibinfo {pages} {061304(R)} (\bibinfo {year} {2012})}\BibitemShut
  {NoStop}%
\bibitem [{\citenamefont {Stroberg}\ \emph {et~al.}(2016)\citenamefont
  {Stroberg}, \citenamefont {Hergert}, \citenamefont {Holt}, \citenamefont
  {Bogner},\ and\ \citenamefont {Schwenk}}]{Stro16TNO}%
  \BibitemOpen
  \bibfield  {author} {\bibinfo {author} {\bibfnamefont {S.~R.}\ \bibnamefont
  {Stroberg}}, \bibinfo {author} {\bibfnamefont {H.}~\bibnamefont {Hergert}},
  \bibinfo {author} {\bibfnamefont {J.~D.}\ \bibnamefont {Holt}}, \bibinfo
  {author} {\bibfnamefont {S.~K.}\ \bibnamefont {Bogner}}, \ and\ \bibinfo
  {author} {\bibfnamefont {A.}~\bibnamefont {Schwenk}},\ }\href {\doibase
  10.1103/PhysRevC.93.051301} {\bibfield  {journal} {\bibinfo  {journal} {Phys.
  Rev. C}\ }\textbf {\bibinfo {volume} {93}},\ \bibinfo {pages} {051301(R)}
  (\bibinfo {year} {2016})}\BibitemShut {NoStop}%
\bibitem [{\citenamefont {Goriely}\ \emph {et~al.}(2009)\citenamefont
  {Goriely}, \citenamefont {Chamel},\ and\ \citenamefont
  {Pearson}}]{Goriely2009}%
  \BibitemOpen
  \bibfield  {author} {\bibinfo {author} {\bibfnamefont {S.}~\bibnamefont
  {Goriely}}, \bibinfo {author} {\bibfnamefont {N.}~\bibnamefont {Chamel}}, \
  and\ \bibinfo {author} {\bibfnamefont {J.~M.}\ \bibnamefont {Pearson}},\
  }\href {\doibase 10.1103/PhysRevLett.102.152503} {\bibfield  {journal}
  {\bibinfo  {journal} {Phys. Rev. Lett.}\ }\textbf {\bibinfo {volume} {102}},\
  \bibinfo {pages} {152503} (\bibinfo {year} {2009})}\BibitemShut {NoStop}%
\bibitem [{\citenamefont {Bulgac}\ \emph {et~al.}(2018)\citenamefont {Bulgac},
  \citenamefont {Forbes}, \citenamefont {Jin}, \citenamefont {Perez},\ and\
  \citenamefont {Schunck}}]{Bulgac2018}%
  \BibitemOpen
  \bibfield  {author} {\bibinfo {author} {\bibfnamefont {A.}~\bibnamefont
  {Bulgac}}, \bibinfo {author} {\bibfnamefont {M.~M.}\ \bibnamefont {Forbes}},
  \bibinfo {author} {\bibfnamefont {S.}~\bibnamefont {Jin}}, \bibinfo {author}
  {\bibfnamefont {R.~N.}\ \bibnamefont {Perez}}, \ and\ \bibinfo {author}
  {\bibfnamefont {N.}~\bibnamefont {Schunck}},\ }\href {\doibase
  10.1103/PhysRevC.97.044313} {\bibfield  {journal} {\bibinfo  {journal} {Phys.
  Rev. C}\ }\textbf {\bibinfo {volume} {97}},\ \bibinfo {pages} {044313}
  (\bibinfo {year} {2018})}\BibitemShut {NoStop}%
\bibitem [{\citenamefont {Navarro~P\'erez}\ \emph {et~al.}(2018)\citenamefont
  {Navarro~P\'erez}, \citenamefont {Schunck}, \citenamefont {Dyhdalo},
  \citenamefont {Furnstahl},\ and\ \citenamefont {Bogner}}]{NavarroPerez2018}%
  \BibitemOpen
  \bibfield  {author} {\bibinfo {author} {\bibfnamefont {R.}~\bibnamefont
  {Navarro~P\'erez}}, \bibinfo {author} {\bibfnamefont {N.}~\bibnamefont
  {Schunck}}, \bibinfo {author} {\bibfnamefont {A.}~\bibnamefont {Dyhdalo}},
  \bibinfo {author} {\bibfnamefont {R.~J.}\ \bibnamefont {Furnstahl}}, \ and\
  \bibinfo {author} {\bibfnamefont {S.~K.}\ \bibnamefont {Bogner}},\ }\href
  {\doibase 10.1103/PhysRevC.97.054304} {\bibfield  {journal} {\bibinfo
  {journal} {Phys. Rev. C}\ }\textbf {\bibinfo {volume} {97}},\ \bibinfo
  {pages} {054304} (\bibinfo {year} {2018})}\BibitemShut {NoStop}%
\bibitem [{\citenamefont {Neufcourt}\ \emph {et~al.}(2018)\citenamefont
  {Neufcourt}, \citenamefont {Cao}, \citenamefont {Nazarewicz},\ and\
  \citenamefont {Viens}}]{Neuf2018}%
  \BibitemOpen
  \bibfield  {author} {\bibinfo {author} {\bibfnamefont {L.}~\bibnamefont
  {Neufcourt}}, \bibinfo {author} {\bibfnamefont {Y.}~\bibnamefont {Cao}},
  \bibinfo {author} {\bibfnamefont {W.}~\bibnamefont {Nazarewicz}}, \ and\
  \bibinfo {author} {\bibfnamefont {F.}~\bibnamefont {Viens}},\ }\href
  {\doibase 10.1103/PhysRevC.98.034318} {\bibfield  {journal} {\bibinfo
  {journal} {Phys. Rev. C}\ }\textbf {\bibinfo {volume} {98}},\ \bibinfo
  {pages} {034318} (\bibinfo {year} {2018})}\BibitemShut {NoStop}%
\bibitem [{\citenamefont {Neufcourt}\ \emph {et~al.}(2020)\citenamefont
  {Neufcourt}, \citenamefont {Cao}, \citenamefont {Giuliani}, \citenamefont
  {Nazarewicz}, \citenamefont {Olsen},\ and\ \citenamefont
  {Tarasov}}]{Neuf2020}%
  \BibitemOpen
  \bibfield  {author} {\bibinfo {author} {\bibfnamefont {L.~L.}\ \bibnamefont
  {Neufcourt}}, \bibinfo {author} {\bibfnamefont {Y.}~\bibnamefont {Cao}},
  \bibinfo {author} {\bibfnamefont {S.~A.}\ \bibnamefont {Giuliani}}, \bibinfo
  {author} {\bibfnamefont {W.}~\bibnamefont {Nazarewicz}}, \bibinfo {author}
  {\bibfnamefont {E.}~\bibnamefont {Olsen}}, \ and\ \bibinfo {author}
  {\bibfnamefont {O.~B.}\ \bibnamefont {Tarasov}},\ }\href {\doibase
  10.1103/PhysRevC.101.044307} {\bibfield  {journal} {\bibinfo  {journal}
  {Phys. Rev. C}\ }\textbf {\bibinfo {volume} {101}},\ \bibinfo {pages}
  {044307} (\bibinfo {year} {2020})}\BibitemShut {NoStop}%
\bibitem [{\citenamefont {Furnstahl}\ \emph {et~al.}(2012)\citenamefont
  {Furnstahl}, \citenamefont {Hagen},\ and\ \citenamefont
  {Papenbrock}}]{Furnstahl2012}%
  \BibitemOpen
  \bibfield  {author} {\bibinfo {author} {\bibfnamefont {R.~J.}\ \bibnamefont
  {Furnstahl}}, \bibinfo {author} {\bibfnamefont {G.}~\bibnamefont {Hagen}}, \
  and\ \bibinfo {author} {\bibfnamefont {T.}~\bibnamefont {Papenbrock}},\
  }\href {\doibase 10.1103/PhysRevC.86.031301} {\bibfield  {journal} {\bibinfo
  {journal} {Phys. Rev. C}\ }\textbf {\bibinfo {volume} {86}},\ \bibinfo
  {pages} {031301} (\bibinfo {year} {2012})}\BibitemShut {NoStop}%
\bibitem [{\citenamefont {Furnstahl}\ \emph {et~al.}(2015)\citenamefont
  {Furnstahl}, \citenamefont {Hagen}, \citenamefont {Papenbrock},\ and\
  \citenamefont {Wendt}}]{Furnstahl2015}%
  \BibitemOpen
  \bibfield  {author} {\bibinfo {author} {\bibfnamefont {R.~J.}\ \bibnamefont
  {Furnstahl}}, \bibinfo {author} {\bibfnamefont {G.}~\bibnamefont {Hagen}},
  \bibinfo {author} {\bibfnamefont {T.}~\bibnamefont {Papenbrock}}, \ and\
  \bibinfo {author} {\bibfnamefont {K.~A.}\ \bibnamefont {Wendt}},\ }\href
  {\doibase 10.1088/0954-3899/42/3/034032} {\bibfield  {journal} {\bibinfo
  {journal} {J. Phys. G}\ }\textbf {\bibinfo {volume} {42}},\ \bibinfo {pages}
  {034032} (\bibinfo {year} {2015})}\BibitemShut {NoStop}%
\bibitem [{\citenamefont {Morris}\ \emph {et~al.}(2015)\citenamefont {Morris},
  \citenamefont {Parzuchowski},\ and\ \citenamefont {Bogner}}]{Morr15Magnus}%
  \BibitemOpen
  \bibfield  {author} {\bibinfo {author} {\bibfnamefont {T.~D.}\ \bibnamefont
  {Morris}}, \bibinfo {author} {\bibfnamefont {N.}~\bibnamefont
  {Parzuchowski}}, \ and\ \bibinfo {author} {\bibfnamefont {S.~K.}\
  \bibnamefont {Bogner}},\ }\href {\doibase 10.1103/PhysRevC.92.034331}
  {\bibfield  {journal} {\bibinfo  {journal} {Phys. Rev. C}\ }\textbf {\bibinfo
  {volume} {92}},\ \bibinfo {pages} {034331} (\bibinfo {year}
  {2015})}\BibitemShut {NoStop}%
\bibitem [{\citenamefont {Brown}\ and\ \citenamefont
  {Rae}(2014)}]{Brow14NuShellX}%
  \BibitemOpen
  \bibfield  {author} {\bibinfo {author} {\bibfnamefont {B.~A.}\ \bibnamefont
  {Brown}}\ and\ \bibinfo {author} {\bibfnamefont {W.~D.~M.}\ \bibnamefont
  {Rae}},\ }\href {\doibase 10.1016/j.nds.2014.07.022} {\bibfield  {journal}
  {\bibinfo  {journal} {Nuclear Data Sheets}\ }\textbf {\bibinfo {volume}
  {120}},\ \bibinfo {pages} {115} (\bibinfo {year} {2014})}\BibitemShut
  {NoStop}%
\bibitem [{\citenamefont {Shimizu}\ \emph {et~al.}(2019)\citenamefont
  {Shimizu}, \citenamefont {Mizusaki}, \citenamefont {Utsuno},\ and\
  \citenamefont {Tsunoda}}]{Shimizu2019}%
  \BibitemOpen
  \bibfield  {author} {\bibinfo {author} {\bibfnamefont {N.}~\bibnamefont
  {Shimizu}}, \bibinfo {author} {\bibfnamefont {T.}~\bibnamefont {Mizusaki}},
  \bibinfo {author} {\bibfnamefont {Y.}~\bibnamefont {Utsuno}}, \ and\ \bibinfo
  {author} {\bibfnamefont {Y.}~\bibnamefont {Tsunoda}},\ }\href {\doibase
  10.1016/j.cpc.2019.06.011} {\bibfield  {journal} {\bibinfo  {journal}
  {Comput. Phys. Commun.}\ }\textbf {\bibinfo {volume} {244}},\ \bibinfo
  {pages} {372} (\bibinfo {year} {2019})}\BibitemShut {NoStop}%
\bibitem [{\citenamefont {Simonis}\ \emph {et~al.}(2016)\citenamefont
  {Simonis}, \citenamefont {Hebeler}, \citenamefont {Holt}, \citenamefont
  {Men{\'e}ndez},\ and\ \citenamefont {Schwenk}}]{Simo16unc}%
  \BibitemOpen
  \bibfield  {author} {\bibinfo {author} {\bibfnamefont {J.}~\bibnamefont
  {Simonis}}, \bibinfo {author} {\bibfnamefont {K.}~\bibnamefont {Hebeler}},
  \bibinfo {author} {\bibfnamefont {J.~D.}\ \bibnamefont {Holt}}, \bibinfo
  {author} {\bibfnamefont {J.}~\bibnamefont {Men{\'e}ndez}}, \ and\ \bibinfo
  {author} {\bibfnamefont {A.}~\bibnamefont {Schwenk}},\ }\href {\doibase
  10.1103/PhysRevC.93.011302} {\bibfield  {journal} {\bibinfo  {journal} {Phys.
  Rev. C}\ }\textbf {\bibinfo {volume} {93}},\ \bibinfo {pages} {011302(R)}
  (\bibinfo {year} {2016})}\BibitemShut {NoStop}%
\bibitem [{\citenamefont {Simonis}\ \emph {et~al.}(2017)\citenamefont
  {Simonis}, \citenamefont {Stroberg}, \citenamefont {Hebeler}, \citenamefont
  {Holt},\ and\ \citenamefont {Schwenk}}]{Simo17SatFinNuc}%
  \BibitemOpen
  \bibfield  {author} {\bibinfo {author} {\bibfnamefont {J.}~\bibnamefont
  {Simonis}}, \bibinfo {author} {\bibfnamefont {S.~R.}\ \bibnamefont
  {Stroberg}}, \bibinfo {author} {\bibfnamefont {K.}~\bibnamefont {Hebeler}},
  \bibinfo {author} {\bibfnamefont {J.~D.}\ \bibnamefont {Holt}}, \ and\
  \bibinfo {author} {\bibfnamefont {A.}~\bibnamefont {Schwenk}},\ }\href
  {\doibase 10.1103/PhysRevC.96.014303} {\bibfield  {journal} {\bibinfo
  {journal} {Phys. Rev. C}\ }\textbf {\bibinfo {volume} {96}},\ \bibinfo
  {pages} {014303} (\bibinfo {year} {2017})}\BibitemShut {NoStop}%
\bibitem [{\citenamefont {Drischler}\ \emph {et~al.}(2019)\citenamefont
  {Drischler}, \citenamefont {Hebeler},\ and\ \citenamefont
  {Schwenk}}]{Dris19nmat}%
  \BibitemOpen
  \bibfield  {author} {\bibinfo {author} {\bibfnamefont {C.}~\bibnamefont
  {Drischler}}, \bibinfo {author} {\bibfnamefont {K.}~\bibnamefont {Hebeler}},
  \ and\ \bibinfo {author} {\bibfnamefont {A.}~\bibnamefont {Schwenk}},\ }\href
  {\doibase 10.1103/PhysRevLett.122.042501} {\bibfield  {journal} {\bibinfo
  {journal} {Phys. Rev. Lett.}\ }\textbf {\bibinfo {volume} {122}},\ \bibinfo
  {pages} {42501} (\bibinfo {year} {2019})}\BibitemShut {NoStop}%
\bibitem [{\citenamefont {Wang}\ \emph {et~al.}(2017)\citenamefont {Wang},
  \citenamefont {Audi}, \citenamefont {Kondev}, \citenamefont {Huang},
  \citenamefont {Naimi},\ and\ \citenamefont {Xu}}]{Wang2017}%
  \BibitemOpen
  \bibfield  {author} {\bibinfo {author} {\bibfnamefont {M.}~\bibnamefont
  {Wang}}, \bibinfo {author} {\bibfnamefont {G.}~\bibnamefont {Audi}}, \bibinfo
  {author} {\bibfnamefont {F.~G.}\ \bibnamefont {Kondev}}, \bibinfo {author}
  {\bibfnamefont {W.~J.}\ \bibnamefont {Huang}}, \bibinfo {author}
  {\bibfnamefont {S.}~\bibnamefont {Naimi}}, \ and\ \bibinfo {author}
  {\bibfnamefont {X.}~\bibnamefont {Xu}},\ }\href
  {http://iopscience.iop.org/article/10.1088/1674-1137/41/3/030003} {\bibfield
  {journal} {\bibinfo  {journal} {Chinese Phys. C}\ }\textbf {\bibinfo {volume}
  {41}},\ \bibinfo {pages} {030003} (\bibinfo {year} {2017})}\BibitemShut
  {NoStop}%
\bibitem [{\citenamefont {Webb}\ \emph {et~al.}(2019)\citenamefont {Webb},
  \citenamefont {Wang}, \citenamefont {Brown}, \citenamefont {Charity},
  \citenamefont {Elson}, \citenamefont {Barney}, \citenamefont {Cerizza},
  \citenamefont {Chajecki}, \citenamefont {Estee}, \citenamefont {Hoff},
  \citenamefont {Kuvin}, \citenamefont {Lynch}, \citenamefont {Manfredi},
  \citenamefont {McNeel}, \citenamefont {Morfouace}, \citenamefont
  {Nazarewicz}, \citenamefont {Pruitt}, \citenamefont {Santamaria},
  \citenamefont {Smith}, \citenamefont {Sobotka}, \citenamefont {Sweany},
  \citenamefont {Tsang}, \citenamefont {Tsang}, \citenamefont {Wuosmaa},
  \citenamefont {Zhang},\ and\ \citenamefont {Zhu}}]{Webb2019}%
  \BibitemOpen
  \bibfield  {author} {\bibinfo {author} {\bibfnamefont {T.~B.}\ \bibnamefont
  {Webb}}, \bibinfo {author} {\bibfnamefont {S.~M.}\ \bibnamefont {Wang}},
  \bibinfo {author} {\bibfnamefont {K.~W.}\ \bibnamefont {Brown}}, \bibinfo
  {author} {\bibfnamefont {R.~J.}\ \bibnamefont {Charity}}, \bibinfo {author}
  {\bibfnamefont {J.~M.}\ \bibnamefont {Elson}}, \bibinfo {author}
  {\bibfnamefont {J.}~\bibnamefont {Barney}}, \bibinfo {author} {\bibfnamefont
  {G.}~\bibnamefont {Cerizza}}, \bibinfo {author} {\bibfnamefont
  {Z.}~\bibnamefont {Chajecki}}, \bibinfo {author} {\bibfnamefont
  {J.}~\bibnamefont {Estee}}, \bibinfo {author} {\bibfnamefont {D.~E.~M.}\
  \bibnamefont {Hoff}}, \bibinfo {author} {\bibfnamefont {S.~A.}\ \bibnamefont
  {Kuvin}}, \bibinfo {author} {\bibfnamefont {W.~G.}\ \bibnamefont {Lynch}},
  \bibinfo {author} {\bibfnamefont {J.}~\bibnamefont {Manfredi}}, \bibinfo
  {author} {\bibfnamefont {D.}~\bibnamefont {McNeel}}, \bibinfo {author}
  {\bibfnamefont {P.}~\bibnamefont {Morfouace}}, \bibinfo {author}
  {\bibfnamefont {W.}~\bibnamefont {Nazarewicz}}, \bibinfo {author}
  {\bibfnamefont {C.~D.}\ \bibnamefont {Pruitt}}, \bibinfo {author}
  {\bibfnamefont {C.}~\bibnamefont {Santamaria}}, \bibinfo {author}
  {\bibfnamefont {J.}~\bibnamefont {Smith}}, \bibinfo {author} {\bibfnamefont
  {L.~G.}\ \bibnamefont {Sobotka}}, \bibinfo {author} {\bibfnamefont
  {S.}~\bibnamefont {Sweany}}, \bibinfo {author} {\bibfnamefont {C.~Y.}\
  \bibnamefont {Tsang}}, \bibinfo {author} {\bibfnamefont {M.~B.}\ \bibnamefont
  {Tsang}}, \bibinfo {author} {\bibfnamefont {A.~H.}\ \bibnamefont {Wuosmaa}},
  \bibinfo {author} {\bibfnamefont {Y.}~\bibnamefont {Zhang}}, \ and\ \bibinfo
  {author} {\bibfnamefont {Z.}~\bibnamefont {Zhu}},\ }\href {\doibase
  10.1103/PhysRevLett.122.122501} {\bibfield  {journal} {\bibinfo  {journal}
  {Phys. Rev. Lett.}\ }\textbf {\bibinfo {volume} {122}},\ \bibinfo {pages}
  {122501} (\bibinfo {year} {2019})}\BibitemShut {NoStop}%
\bibitem [{\citenamefont {Leblond}\ \emph {et~al.}(2018)\citenamefont
  {Leblond}, \citenamefont {Marqu{\'{e}}s}, \citenamefont {Gibelin},
  \citenamefont {Orr}, \citenamefont {Kondo}, \citenamefont {Nakamura},
  \citenamefont {Bonnard}, \citenamefont {Michel}, \citenamefont {Achouri},
  \citenamefont {Aumann}, \citenamefont {Baba}, \citenamefont {Delaunay},
  \citenamefont {Deshayes}, \citenamefont {Doornenbal}, \citenamefont {Fukuda},
  \citenamefont {Hwang}, \citenamefont {Inabe}, \citenamefont {Isobe},
  \citenamefont {Kameda}, \citenamefont {Kanno}, \citenamefont {Kim},
  \citenamefont {Kobayashi}, \citenamefont {Kobayashi}, \citenamefont {Kubo},
  \citenamefont {Lee}, \citenamefont {Minakata}, \citenamefont {Motobayashi},
  \citenamefont {Murai}, \citenamefont {Murakami}, \citenamefont {Muto},
  \citenamefont {Nakashima}, \citenamefont {Nakatsuka}, \citenamefont {Navin},
  \citenamefont {Nishi}, \citenamefont {Ogoshi}, \citenamefont {Otsu},
  \citenamefont {Sato}, \citenamefont {Satou}, \citenamefont {Shimizu},
  \citenamefont {Suzuki}, \citenamefont {Takahashi}, \citenamefont {Takeda},
  \citenamefont {Takeuchi}, \citenamefont {Tanaka}, \citenamefont {Togano},
  \citenamefont {Tuff}, \citenamefont {Vandebrouck},\ and\ \citenamefont
  {Yoneda}}]{Leblond2018}%
  \BibitemOpen
  \bibfield  {author} {\bibinfo {author} {\bibfnamefont {S.}~\bibnamefont
  {Leblond}}, \bibinfo {author} {\bibfnamefont {F.~M.}\ \bibnamefont
  {Marqu{\'{e}}s}}, \bibinfo {author} {\bibfnamefont {J.}~\bibnamefont
  {Gibelin}}, \bibinfo {author} {\bibfnamefont {N.~A.}\ \bibnamefont {Orr}},
  \bibinfo {author} {\bibfnamefont {Y.}~\bibnamefont {Kondo}}, \bibinfo
  {author} {\bibfnamefont {T.}~\bibnamefont {Nakamura}}, \bibinfo {author}
  {\bibfnamefont {J.}~\bibnamefont {Bonnard}}, \bibinfo {author} {\bibfnamefont
  {N.}~\bibnamefont {Michel}}, \bibinfo {author} {\bibfnamefont {N.~L.}\
  \bibnamefont {Achouri}}, \bibinfo {author} {\bibfnamefont {T.}~\bibnamefont
  {Aumann}}, \bibinfo {author} {\bibfnamefont {H.}~\bibnamefont {Baba}},
  \bibinfo {author} {\bibfnamefont {F.}~\bibnamefont {Delaunay}}, \bibinfo
  {author} {\bibfnamefont {Q.}~\bibnamefont {Deshayes}}, \bibinfo {author}
  {\bibfnamefont {P.}~\bibnamefont {Doornenbal}}, \bibinfo {author}
  {\bibfnamefont {N.}~\bibnamefont {Fukuda}}, \bibinfo {author} {\bibfnamefont
  {J.~W.}\ \bibnamefont {Hwang}}, \bibinfo {author} {\bibfnamefont
  {N.}~\bibnamefont {Inabe}}, \bibinfo {author} {\bibfnamefont
  {T.}~\bibnamefont {Isobe}}, \bibinfo {author} {\bibfnamefont
  {D.}~\bibnamefont {Kameda}}, \bibinfo {author} {\bibfnamefont
  {D.}~\bibnamefont {Kanno}}, \bibinfo {author} {\bibfnamefont
  {S.}~\bibnamefont {Kim}}, \bibinfo {author} {\bibfnamefont {N.}~\bibnamefont
  {Kobayashi}}, \bibinfo {author} {\bibfnamefont {T.}~\bibnamefont
  {Kobayashi}}, \bibinfo {author} {\bibfnamefont {T.}~\bibnamefont {Kubo}},
  \bibinfo {author} {\bibfnamefont {J.}~\bibnamefont {Lee}}, \bibinfo {author}
  {\bibfnamefont {R.}~\bibnamefont {Minakata}}, \bibinfo {author}
  {\bibfnamefont {T.}~\bibnamefont {Motobayashi}}, \bibinfo {author}
  {\bibfnamefont {D.}~\bibnamefont {Murai}}, \bibinfo {author} {\bibfnamefont
  {T.}~\bibnamefont {Murakami}}, \bibinfo {author} {\bibfnamefont
  {K.}~\bibnamefont {Muto}}, \bibinfo {author} {\bibfnamefont {T.}~\bibnamefont
  {Nakashima}}, \bibinfo {author} {\bibfnamefont {N.}~\bibnamefont
  {Nakatsuka}}, \bibinfo {author} {\bibfnamefont {A.}~\bibnamefont {Navin}},
  \bibinfo {author} {\bibfnamefont {S.}~\bibnamefont {Nishi}}, \bibinfo
  {author} {\bibfnamefont {S.}~\bibnamefont {Ogoshi}}, \bibinfo {author}
  {\bibfnamefont {H.}~\bibnamefont {Otsu}}, \bibinfo {author} {\bibfnamefont
  {H.}~\bibnamefont {Sato}}, \bibinfo {author} {\bibfnamefont {Y.}~\bibnamefont
  {Satou}}, \bibinfo {author} {\bibfnamefont {Y.}~\bibnamefont {Shimizu}},
  \bibinfo {author} {\bibfnamefont {H.}~\bibnamefont {Suzuki}}, \bibinfo
  {author} {\bibfnamefont {K.}~\bibnamefont {Takahashi}}, \bibinfo {author}
  {\bibfnamefont {H.}~\bibnamefont {Takeda}}, \bibinfo {author} {\bibfnamefont
  {S.}~\bibnamefont {Takeuchi}}, \bibinfo {author} {\bibfnamefont
  {R.}~\bibnamefont {Tanaka}}, \bibinfo {author} {\bibfnamefont
  {Y.}~\bibnamefont {Togano}}, \bibinfo {author} {\bibfnamefont {A.~G.}\
  \bibnamefont {Tuff}}, \bibinfo {author} {\bibfnamefont {M.}~\bibnamefont
  {Vandebrouck}}, \ and\ \bibinfo {author} {\bibfnamefont {K.}~\bibnamefont
  {Yoneda}},\ }\href {\doibase 10.1103/PhysRevLett.121.262502} {\bibfield
  {journal} {\bibinfo  {journal} {Phys. Rev. Lett.}\ }\textbf {\bibinfo
  {volume} {121}},\ \bibinfo {pages} {262502} (\bibinfo {year}
  {2018})}\BibitemShut {NoStop}%
\bibitem [{\citenamefont {Mukha}\ \emph {et~al.}(2018)\citenamefont {Mukha},
  \citenamefont {Grigorenko}, \citenamefont {Kostyleva}, \citenamefont
  {Acosta}, \citenamefont {Casarejos}, \citenamefont {Ciemny}, \citenamefont
  {Dominik}, \citenamefont {Due{\~{n}}as}, \citenamefont {Dunin}, \citenamefont
  {Espino}, \citenamefont {Estrad{\'{e}}}, \citenamefont {Farinon},
  \citenamefont {Fomichev}, \citenamefont {Geissel}, \citenamefont {Gorshkov},
  \citenamefont {Janas}, \citenamefont {Kami{\'{n}}ski}, \citenamefont
  {Kiselev}, \citenamefont {Kn{\"{o}}bel}, \citenamefont {Krupko},
  \citenamefont {Kuich}, \citenamefont {Litvinov}, \citenamefont
  {Marquinez-Dur{\'{a}}n}, \citenamefont {Martel}, \citenamefont {Mazzocchi},
  \citenamefont {Nociforo}, \citenamefont {Ord{\'{u}}z}, \citenamefont
  {Pf{\"{u}}tzner}, \citenamefont {Pietri}, \citenamefont {Pomorski},
  \citenamefont {Prochazka}, \citenamefont {Rymzhanova}, \citenamefont
  {S{\'{a}}nchez-Ben{\'{i}}tez}, \citenamefont {Scheidenberger}, \citenamefont
  {Sharov}, \citenamefont {Simon}, \citenamefont {Sitar}, \citenamefont
  {Slepnev}, \citenamefont {Stanoiu}, \citenamefont {Strmen}, \citenamefont
  {Szarka}, \citenamefont {Takechi}, \citenamefont {Tanaka}, \citenamefont
  {Weick}, \citenamefont {Winkler}, \citenamefont {Winfield}, \citenamefont
  {Xu},\ and\ \citenamefont {Zhukov}}]{Mukha2018}%
  \BibitemOpen
  \bibfield  {author} {\bibinfo {author} {\bibfnamefont {I.}~\bibnamefont
  {Mukha}}, \bibinfo {author} {\bibfnamefont {L.~V.}\ \bibnamefont
  {Grigorenko}}, \bibinfo {author} {\bibfnamefont {D.}~\bibnamefont
  {Kostyleva}}, \bibinfo {author} {\bibfnamefont {L.}~\bibnamefont {Acosta}},
  \bibinfo {author} {\bibfnamefont {E.}~\bibnamefont {Casarejos}}, \bibinfo
  {author} {\bibfnamefont {A.~A.}\ \bibnamefont {Ciemny}}, \bibinfo {author}
  {\bibfnamefont {W.}~\bibnamefont {Dominik}}, \bibinfo {author} {\bibfnamefont
  {J.~A.}\ \bibnamefont {Due{\~{n}}as}}, \bibinfo {author} {\bibfnamefont
  {V.}~\bibnamefont {Dunin}}, \bibinfo {author} {\bibfnamefont {J.~M.}\
  \bibnamefont {Espino}}, \bibinfo {author} {\bibfnamefont {A.}~\bibnamefont
  {Estrad{\'{e}}}}, \bibinfo {author} {\bibfnamefont {F.}~\bibnamefont
  {Farinon}}, \bibinfo {author} {\bibfnamefont {A.}~\bibnamefont {Fomichev}},
  \bibinfo {author} {\bibfnamefont {H.}~\bibnamefont {Geissel}}, \bibinfo
  {author} {\bibfnamefont {A.}~\bibnamefont {Gorshkov}}, \bibinfo {author}
  {\bibfnamefont {Z.}~\bibnamefont {Janas}}, \bibinfo {author} {\bibfnamefont
  {G.}~\bibnamefont {Kami{\'{n}}ski}}, \bibinfo {author} {\bibfnamefont
  {O.}~\bibnamefont {Kiselev}}, \bibinfo {author} {\bibfnamefont
  {R.}~\bibnamefont {Kn{\"{o}}bel}}, \bibinfo {author} {\bibfnamefont
  {S.}~\bibnamefont {Krupko}}, \bibinfo {author} {\bibfnamefont
  {M.}~\bibnamefont {Kuich}}, \bibinfo {author} {\bibfnamefont {Y.~A.}\
  \bibnamefont {Litvinov}}, \bibinfo {author} {\bibfnamefont {G.}~\bibnamefont
  {Marquinez-Dur{\'{a}}n}}, \bibinfo {author} {\bibfnamefont {I.}~\bibnamefont
  {Martel}}, \bibinfo {author} {\bibfnamefont {C.}~\bibnamefont {Mazzocchi}},
  \bibinfo {author} {\bibfnamefont {C.}~\bibnamefont {Nociforo}}, \bibinfo
  {author} {\bibfnamefont {A.~K.}\ \bibnamefont {Ord{\'{u}}z}}, \bibinfo
  {author} {\bibfnamefont {M.}~\bibnamefont {Pf{\"{u}}tzner}}, \bibinfo
  {author} {\bibfnamefont {S.}~\bibnamefont {Pietri}}, \bibinfo {author}
  {\bibfnamefont {M.}~\bibnamefont {Pomorski}}, \bibinfo {author}
  {\bibfnamefont {A.}~\bibnamefont {Prochazka}}, \bibinfo {author}
  {\bibfnamefont {S.}~\bibnamefont {Rymzhanova}}, \bibinfo {author}
  {\bibfnamefont {A.~M.}\ \bibnamefont {S{\'{a}}nchez-Ben{\'{i}}tez}}, \bibinfo
  {author} {\bibfnamefont {C.}~\bibnamefont {Scheidenberger}}, \bibinfo
  {author} {\bibfnamefont {P.}~\bibnamefont {Sharov}}, \bibinfo {author}
  {\bibfnamefont {H.}~\bibnamefont {Simon}}, \bibinfo {author} {\bibfnamefont
  {B.}~\bibnamefont {Sitar}}, \bibinfo {author} {\bibfnamefont
  {R.}~\bibnamefont {Slepnev}}, \bibinfo {author} {\bibfnamefont
  {M.}~\bibnamefont {Stanoiu}}, \bibinfo {author} {\bibfnamefont
  {P.}~\bibnamefont {Strmen}}, \bibinfo {author} {\bibfnamefont
  {I.}~\bibnamefont {Szarka}}, \bibinfo {author} {\bibfnamefont
  {M.}~\bibnamefont {Takechi}}, \bibinfo {author} {\bibfnamefont {Y.~K.}\
  \bibnamefont {Tanaka}}, \bibinfo {author} {\bibfnamefont {H.}~\bibnamefont
  {Weick}}, \bibinfo {author} {\bibfnamefont {M.}~\bibnamefont {Winkler}},
  \bibinfo {author} {\bibfnamefont {J.~S.}\ \bibnamefont {Winfield}}, \bibinfo
  {author} {\bibfnamefont {X.}~\bibnamefont {Xu}}, \ and\ \bibinfo {author}
  {\bibfnamefont {M.~V.}\ \bibnamefont {Zhukov}},\ }\href {\doibase
  10.1103/PhysRevC.98.064308} {\bibfield  {journal} {\bibinfo  {journal} {Phys.
  Rev. C}\ }\textbf {\bibinfo {volume} {98}},\ \bibinfo {pages} {064308}
  (\bibinfo {year} {2018})}\BibitemShut {NoStop}%
\bibitem [{\citenamefont {Michimasa}\ \emph {et~al.}(2018)\citenamefont
  {Michimasa}, \citenamefont {Kobayashi}, \citenamefont {Kiyokawa},
  \citenamefont {Ota}, \citenamefont {Ahn}, \citenamefont {Baba}, \citenamefont
  {Berg}, \citenamefont {Dozono}, \citenamefont {Fukuda}, \citenamefont
  {Furuno}, \citenamefont {Ideguchi}, \citenamefont {Inabe}, \citenamefont
  {Kawabata}, \citenamefont {Kawase}, \citenamefont {Kisamori}, \citenamefont
  {Kobayashi}, \citenamefont {Kubo}, \citenamefont {Kubota}, \citenamefont
  {Lee}, \citenamefont {Matsushita}, \citenamefont {Miya}, \citenamefont
  {Mizukami}, \citenamefont {Nagakura}, \citenamefont {Nishimura},
  \citenamefont {Oikawa}, \citenamefont {Sakai}, \citenamefont {Shimizu},
  \citenamefont {Stolz}, \citenamefont {Suzuki}, \citenamefont {Takaki},
  \citenamefont {Takeda}, \citenamefont {Takeuchi}, \citenamefont {Tokieda},
  \citenamefont {Uesaka}, \citenamefont {Yako}, \citenamefont {Yamaguchi},
  \citenamefont {Yanagisawa}, \citenamefont {Yokoyama}, \citenamefont
  {Yoshida},\ and\ \citenamefont {Shimoura}}]{Michimasa2018}%
  \BibitemOpen
  \bibfield  {author} {\bibinfo {author} {\bibfnamefont {S.}~\bibnamefont
  {Michimasa}}, \bibinfo {author} {\bibfnamefont {M.}~\bibnamefont
  {Kobayashi}}, \bibinfo {author} {\bibfnamefont {Y.}~\bibnamefont {Kiyokawa}},
  \bibinfo {author} {\bibfnamefont {S.}~\bibnamefont {Ota}}, \bibinfo {author}
  {\bibfnamefont {D.~S.}\ \bibnamefont {Ahn}}, \bibinfo {author} {\bibfnamefont
  {H.}~\bibnamefont {Baba}}, \bibinfo {author} {\bibfnamefont {G.~P.~A.}\
  \bibnamefont {Berg}}, \bibinfo {author} {\bibfnamefont {M.}~\bibnamefont
  {Dozono}}, \bibinfo {author} {\bibfnamefont {N.}~\bibnamefont {Fukuda}},
  \bibinfo {author} {\bibfnamefont {T.}~\bibnamefont {Furuno}}, \bibinfo
  {author} {\bibfnamefont {E.}~\bibnamefont {Ideguchi}}, \bibinfo {author}
  {\bibfnamefont {N.}~\bibnamefont {Inabe}}, \bibinfo {author} {\bibfnamefont
  {T.}~\bibnamefont {Kawabata}}, \bibinfo {author} {\bibfnamefont
  {S.}~\bibnamefont {Kawase}}, \bibinfo {author} {\bibfnamefont
  {K.}~\bibnamefont {Kisamori}}, \bibinfo {author} {\bibfnamefont
  {K.}~\bibnamefont {Kobayashi}}, \bibinfo {author} {\bibfnamefont
  {T.}~\bibnamefont {Kubo}}, \bibinfo {author} {\bibfnamefont {Y.}~\bibnamefont
  {Kubota}}, \bibinfo {author} {\bibfnamefont {C.~S.}\ \bibnamefont {Lee}},
  \bibinfo {author} {\bibfnamefont {M.}~\bibnamefont {Matsushita}}, \bibinfo
  {author} {\bibfnamefont {H.}~\bibnamefont {Miya}}, \bibinfo {author}
  {\bibfnamefont {A.}~\bibnamefont {Mizukami}}, \bibinfo {author}
  {\bibfnamefont {H.}~\bibnamefont {Nagakura}}, \bibinfo {author}
  {\bibfnamefont {D.}~\bibnamefont {Nishimura}}, \bibinfo {author}
  {\bibfnamefont {H.}~\bibnamefont {Oikawa}}, \bibinfo {author} {\bibfnamefont
  {H.}~\bibnamefont {Sakai}}, \bibinfo {author} {\bibfnamefont
  {Y.}~\bibnamefont {Shimizu}}, \bibinfo {author} {\bibfnamefont
  {A.}~\bibnamefont {Stolz}}, \bibinfo {author} {\bibfnamefont
  {H.}~\bibnamefont {Suzuki}}, \bibinfo {author} {\bibfnamefont
  {M.}~\bibnamefont {Takaki}}, \bibinfo {author} {\bibfnamefont
  {H.}~\bibnamefont {Takeda}}, \bibinfo {author} {\bibfnamefont
  {S.}~\bibnamefont {Takeuchi}}, \bibinfo {author} {\bibfnamefont
  {H.}~\bibnamefont {Tokieda}}, \bibinfo {author} {\bibfnamefont
  {T.}~\bibnamefont {Uesaka}}, \bibinfo {author} {\bibfnamefont
  {K.}~\bibnamefont {Yako}}, \bibinfo {author} {\bibfnamefont {Y.}~\bibnamefont
  {Yamaguchi}}, \bibinfo {author} {\bibfnamefont {Y.}~\bibnamefont
  {Yanagisawa}}, \bibinfo {author} {\bibfnamefont {R.}~\bibnamefont
  {Yokoyama}}, \bibinfo {author} {\bibfnamefont {K.}~\bibnamefont {Yoshida}}, \
  and\ \bibinfo {author} {\bibfnamefont {S.}~\bibnamefont {Shimoura}},\ }\href
  {\doibase 10.1103/PhysRevLett.121.022506} {\bibfield  {journal} {\bibinfo
  {journal} {Phys. Rev. Lett.}\ }\textbf {\bibinfo {volume} {121}},\ \bibinfo
  {pages} {022506} (\bibinfo {year} {2018})}\BibitemShut {NoStop}%
\bibitem [{\citenamefont {Warburton}\ \emph {et~al.}(1990)\citenamefont
  {Warburton}, \citenamefont {Becker},\ and\ \citenamefont
  {Brown}}]{Warburton1990a}%
  \BibitemOpen
  \bibfield  {author} {\bibinfo {author} {\bibfnamefont {E.~K.}\ \bibnamefont
  {Warburton}}, \bibinfo {author} {\bibfnamefont {J.~A.}\ \bibnamefont
  {Becker}}, \ and\ \bibinfo {author} {\bibfnamefont {B.~A.}\ \bibnamefont
  {Brown}},\ }\href {\doibase 10.1103/PhysRevC.41.1147} {\bibfield  {journal}
  {\bibinfo  {journal} {Phys. Rev. C}\ }\textbf {\bibinfo {volume} {41}},\
  \bibinfo {pages} {1147} (\bibinfo {year} {1990})}\BibitemShut {NoStop}%
\bibitem [{\citenamefont {Caurier}\ \emph {et~al.}(2014)\citenamefont
  {Caurier}, \citenamefont {Nowacki},\ and\ \citenamefont
  {Poves}}]{Caur14n20n28}%
  \BibitemOpen
  \bibfield  {author} {\bibinfo {author} {\bibfnamefont {E.}~\bibnamefont
  {Caurier}}, \bibinfo {author} {\bibfnamefont {F.}~\bibnamefont {Nowacki}}, \
  and\ \bibinfo {author} {\bibfnamefont {A.}~\bibnamefont {Poves}},\ }\href
  {\doibase 10.1103/PhysRevC.90.014302} {\bibfield  {journal} {\bibinfo
  {journal} {Phys. Rev. C}\ }\textbf {\bibinfo {volume} {90}},\ \bibinfo
  {pages} {014302} (\bibinfo {year} {2014})}\BibitemShut {NoStop}%
\bibitem [{\citenamefont {Hagen}\ \emph {et~al.}(2013)\citenamefont {Hagen},
  \citenamefont {Hagen}, \citenamefont {Hammer},\ and\ \citenamefont
  {Platter}}]{Hagen2013}%
  \BibitemOpen
  \bibfield  {author} {\bibinfo {author} {\bibfnamefont {G.}~\bibnamefont
  {Hagen}}, \bibinfo {author} {\bibfnamefont {P.}~\bibnamefont {Hagen}},
  \bibinfo {author} {\bibfnamefont {H.~W.}\ \bibnamefont {Hammer}}, \ and\
  \bibinfo {author} {\bibfnamefont {L.}~\bibnamefont {Platter}},\ }\href
  {\doibase 10.1103/PhysRevLett.111.132501} {\bibfield  {journal} {\bibinfo
  {journal} {Phys. Rev. Lett.}\ }\textbf {\bibinfo {volume} {111}},\ \bibinfo
  {pages} {132501} (\bibinfo {year} {2013})}\BibitemShut {NoStop}%
\bibitem [{\citenamefont {Holt}\ \emph {et~al.}(2014)\citenamefont {Holt},
  \citenamefont {Men{\'{e}}ndez}, \citenamefont {Simonis},\ and\ \citenamefont
  {Schwenk}}]{Holt2014a}%
  \BibitemOpen
  \bibfield  {author} {\bibinfo {author} {\bibfnamefont {J.~D.}\ \bibnamefont
  {Holt}}, \bibinfo {author} {\bibfnamefont {J.}~\bibnamefont
  {Men{\'{e}}ndez}}, \bibinfo {author} {\bibfnamefont {J.}~\bibnamefont
  {Simonis}}, \ and\ \bibinfo {author} {\bibfnamefont {A.}~\bibnamefont
  {Schwenk}},\ }\href {\doibase 10.1103/PhysRevC.90.024312} {\bibfield
  {journal} {\bibinfo  {journal} {Phys. Rev. C}\ }\textbf {\bibinfo {volume}
  {90}},\ \bibinfo {pages} {024312} (\bibinfo {year} {2014})}\BibitemShut
  {NoStop}%
\bibitem [{\citenamefont {Hergert}(2020)}]{Hergert2020}%
  \BibitemOpen
  \bibfield  {author} {\bibinfo {author} {\bibfnamefont {H.}~\bibnamefont
  {Hergert}},\ }\href {\doibase 10.3389/fphy.2020.00379} {\bibfield  {journal}
  {\bibinfo  {journal} {Front. Phys.}\ }\textbf {\bibinfo {volume} {8}},\
  \bibinfo {pages} {1} (\bibinfo {year} {2020})},\ \Eprint
  {http://arxiv.org/abs/2008.05061} {arXiv:2008.05061} \BibitemShut {NoStop}%
\bibitem [{\citenamefont {Gelman}\ \emph {et~al.}(2014)\citenamefont {Gelman},
  \citenamefont {Carlin}, \citenamefont {Stern}, \citenamefont {Dunson},
  \citenamefont {Vehtari},\ and\ \citenamefont {Rubin}}]{Gelman2014}%
  \BibitemOpen
  \bibfield  {author} {\bibinfo {author} {\bibfnamefont {A.}~\bibnamefont
  {Gelman}}, \bibinfo {author} {\bibfnamefont {J.~B.}\ \bibnamefont {Carlin}},
  \bibinfo {author} {\bibfnamefont {H.~S.}\ \bibnamefont {Stern}}, \bibinfo
  {author} {\bibfnamefont {D.~B.}\ \bibnamefont {Dunson}}, \bibinfo {author}
  {\bibfnamefont {A.}~\bibnamefont {Vehtari}}, \ and\ \bibinfo {author}
  {\bibfnamefont {D.~B.}\ \bibnamefont {Rubin}},\ }\href@noop {} {\emph
  {\bibinfo {title} {{Bayesian Data Analysis}}}},\ \bibinfo {edition} {3rd}\
  ed.\ (\bibinfo  {publisher} {CRC Press},\ \bibinfo {address} {Boca Raton},\
  \bibinfo {year} {2014})\BibitemShut {NoStop}%
\bibitem [{\citenamefont {Fahrmeir}\ \emph {et~al.}(2009)\citenamefont
  {Fahrmeir}, \citenamefont {Kneib},\ and\ \citenamefont
  {Lang}}]{Fahrmeir2009}%
  \BibitemOpen
  \bibfield  {author} {\bibinfo {author} {\bibfnamefont {L.}~\bibnamefont
  {Fahrmeir}}, \bibinfo {author} {\bibfnamefont {T.}~\bibnamefont {Kneib}}, \
  and\ \bibinfo {author} {\bibfnamefont {S.}~\bibnamefont {Lang}},\ }\href
  {\doibase 10.1007/978-3-642-01837-4} {\emph {\bibinfo {title}
  {{Regression}}}}\ (\bibinfo  {publisher} {Springer Berlin Heidelberg},\
  \bibinfo {address} {Berlin, Heidelberg},\ \bibinfo {year} {2009})\BibitemShut
  {NoStop}%
\bibitem [{\citenamefont {Newcombe}(1998)}]{Newcombe1998}%
  \BibitemOpen
  \bibfield  {author} {\bibinfo {author} {\bibfnamefont {R.~G.}\ \bibnamefont
  {Newcombe}},\ }\href {\doibase
  10.1002/(SICI)1097-0258(19980430)17:8<857::AID-SIM777>3.0.CO;2-E} {\bibfield
  {journal} {\bibinfo  {journal} {Stat. Med.}\ }\textbf {\bibinfo {volume}
  {17}},\ \bibinfo {pages} {857} (\bibinfo {year} {1998})}\BibitemShut
  {NoStop}%
\end{thebibliography}%
	
	\clearpage
	
	\section{Supplemental Material}
	
	\subsection{Bayesian linear regression}
	
	Our assessment of the residuals $\delta S$ employs a Bayesian linear regression with conjugate priors (see, e.g., Refs.~\cite{Gelman2014,Fahrmeir2009}).
	Our linear model is
	\begin{equation}
	\delta S_{i,\alpha} = \mathcal{A}_{\alpha} S_{i,\alpha}^{\rm th} + \mathcal{B}_{\alpha} + \epsilon_{i,\alpha}(\sigma^2_{\alpha}) \,.
	\end{equation}
	The index $i$ labels a specific nuclide, and the index $\alpha\in\{ n,p,2n,2p\}$ labels a particular separation energy channel.
	The coefficients $\mathcal{A}_{\alpha}$, $\mathcal{B}_{\alpha}$, and the variances $\sigma^2_{\alpha}$ are unknown parameters with distributions to be determined.
	The term $\epsilon_{i,\alpha}(\sigma^2_{\alpha})$ denotes a random number drawn from a Gaussian distribution of mean zero and variance $\sigma^2_{\alpha}$.
	We collect $\mathcal{A}_{\alpha}$ and $\mathcal{B}_{\alpha}$ into a two-dimensional column vector $\bm{\beta}_{\alpha} = \begin{pmatrix} \mathcal{A}_{\alpha} &\mathcal{B}_{\alpha}\end{pmatrix}^{T}$.
	Given $m$ data points, the residuals for a given channel $\alpha$ form an $m$-dimensional column vector
	\begin{equation}
	\bm{y}_{\alpha} = \begin{pmatrix}\delta S_{1,\alpha}& \delta S_{2,\alpha}&\ldots&\delta S_{m,\alpha}\end{pmatrix}^{T} \,,
	\end{equation}
	and the {\it ab initio} separation energies $S^{\rm th}_{i,\alpha}$ are collected in an $m\times 2$ matrix
	\begin{equation}
	X_{\alpha} = \begin{pmatrix} S^{\rm th}_{1,\alpha} & S^{\rm th}_{2,\alpha} & \ldots & S^{\rm th}_{m,\alpha} \\ 1 & 1 & \ldots & 1
	\end{pmatrix}^T \,.
	\end{equation}
	Our linear model becomes
	\begin{equation}
	\bm{y}_{\alpha} = X_{\alpha} \bm{\beta}_{\alpha} + \bm{\epsilon}_{\alpha}(\sigma^2_{\alpha}) \,,
	\end{equation}
	where the bold symbol $\bm{\epsilon}_{\alpha}$ indicates a column vector of $m$ random numbers drawn from a Gaussian of variance $\sigma^2_{\alpha}$.
	
	In the following, we focus on a single channel and suppress the label $\alpha$ in order to avoid cluttering the notation.
	Our priors for $\bm{\beta},\sigma^2$ are taken to be a normal-inverse-gamma distribution
	\begin{equation}
	p(\bm{\beta},\sigma^2) = p(\sigma^2) p(\bm{\beta} | \sigma^2) \,,
	\end{equation}
	with $p(\sigma^2)$ given by an inverse-gamma distribution
	\begin{equation}
	p(\sigma^2) = \frac{b_0^{a_0}}{\Gamma(a_0)} (\sigma^2)^{-a_0-1} \exp(-b_0/\sigma^2) \,,
	\end{equation}
	where $\Gamma$ is the gamma function, and $p(\beta|\sigma^2)$ a multivariate normal distribution
	\begin{equation}
	p(\bm{\beta}|\sigma^2) \propto \frac{1}{\sigma^2} \exp \left[-\frac{ (\bm{\beta} - \bm{\mu_0})^{T} \Lambda_0 (\bm{\beta} - \bm{\mu_0})}{2\sigma^2} \right].
	\end{equation}
	The parameters $a_0,b_0,\bm{\mu_0},\Lambda_0$ characterize our priors.
	We choose relatively noninformative values of $a_0=b_0=1$, $\bm{\mu_0}=\begin{pmatrix}0&0\end{pmatrix}^T$, and $\Lambda_0 = \begin{pmatrix} 0 & 0 \\ 0 & 0\end{pmatrix}$.
	The likelihood function for $\bm{y}$ given $\bm{\beta},\sigma^2,X$ is a multivariate Gaussian
	\begin{equation}
	\begin{aligned}
	p(\bm{y}|\bm{\beta},\sigma^2,X) =& \\ \left(\frac{1}{2\pi\sigma^2}\right)^{m/2} &\exp\left[-\frac{1}{2\sigma^2} (\bm{y}-X\bm{\beta})^T(\bm{y}-X\bm{\beta})\right] .
	\end{aligned}
	\end{equation}
	The posterior distribution for the parameters $\bm{\beta},\sigma^2$ is obtained with Bayes' theorem
	\begin{equation}
	p(\bm{\beta},\sigma^2|\bm{y},X) \propto p(\bm{\beta},\sigma^2)p(\bm{y}|\bm{\beta},\sigma^2,X) \,.
	\end{equation}
	Due to our choice of conjugate priors, the posterior distribution has the same functional form as our priors, but with updated parameters~\cite{Fahrmeir2009}, which we indicate with a subscript $*$:
	\begin{equation}
	\begin{aligned}
	&\Lambda_* = \Lambda_0 +X^{T} X \\
	&\bm{\mu_*} = \Lambda_*^{-1}(\Lambda_0 \bm{\mu_0} + X^T y) \\
	&a_* = a_0 + \frac{m}{2} \\
	&b_* = b_0 + \frac{1}{2}(\bm{\mu_0}^T \Lambda_0 \bm{\mu_0} + y^T y - \mu_*^T \Lambda_* \mu_* ) \,.
	\end{aligned}
	\end{equation}
	The posterior predictive distribution (PPD) for new data $\bm{\tilde{y}}$ given the training data $\bm{y},X$ and corresponding {\it ab initio} calculations $\tilde{X}$ is
	\begin{equation}
	p(\bm{\tilde{y}}|\bm{y},X,\tilde{X}) = \int d\bm{\beta}d\sigma^2 p(\bm{\tilde{y}} | \bm{\beta},\sigma^2,\tilde{X} ) p(\bm{\beta},\sigma^2|\bm{y},X) \,.
	\end{equation}
	This integral can be evaluated analytically, resulting in a multivariate Student-t distribution with $2a_{*}$ degrees of freedom.
	For many degrees of freedom, the Student-t distribution approaches a normal distribution, and given that in the present case $m>100$, we take the PPD to be a multivariate normal.
	The PPD for new data $\bm{\tilde{y}}$, given the training data $\bm{y},X$, and additional {\it ab initio} calculations $\tilde{X}$ is therefore
	\begin{equation}
	p(\bm{\tilde{y}} | \bm{y},X,\tilde{X}) \approx \mathcal{N}( \bm{\tilde{M}}, \tilde{\Sigma} ) 
	\end{equation}
	with mean vector
	\begin{equation}
	\bm{\tilde{M}} = \tilde{X}\bm{\mu_*}
	\end{equation}
	and covariance matrix
	\begin{equation}
	\tilde{\Sigma} = \frac{b_*}{a_*} (\mathbb{I} + \tilde{X} \Lambda_*^{-1} \tilde{X}^T).
	\end{equation}
	
	The above formulation does not account for experimental error bars and, particularly near the drip lines, experimental errors can be comparable to the theoretical uncertainty.
	To preserve the advantages of conjugate priors, we incorporate the experimental uncertainties by Monte Carlo sampling the experimental data, interpreting the reported error bars as indicating one standard deviation of a normal distribution.
	The parameters $\Lambda_*$ and $a_*$ are unaffected by this procedure, since they do not depend on $\bm{y}$,
	but we obtain distributions for $b_*$ and $\bm{\mu_*}$.
	These distributions are unimodal and relatively sharply peaked.
	We have confirmed that taking the mean values of these distributions yields a PPD indistinguishable from the PPD obtained by marginalizing over their distributions.
	(The resulting PPD is, however, slightly broader than what would be obtained by neglecting the experimental uncertainties altogether).
	
	\subsection{Correlation of residuals\\ between neighboring nuclides}
	
\begin{figure}
		\centering
		\includegraphics[width=1.0\columnwidth]{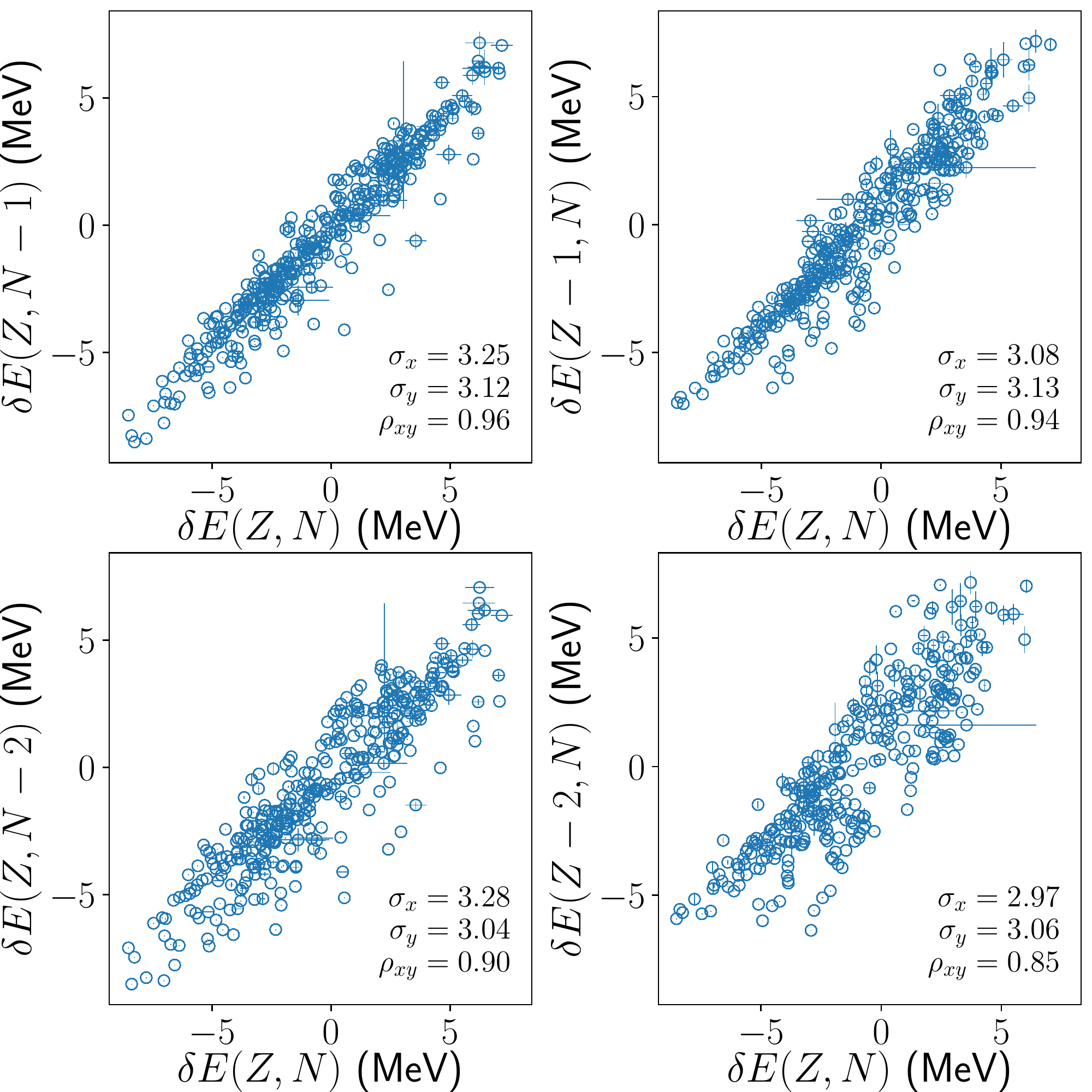}
		\caption{Correlation of the ground-state energy residuals $\delta E$ for neighboring nuclei. In each panel, $\sigma_x,\sigma_y$ indicate the standard deviation along each axis (in MeV) and $\rho_{xy}$ is the dimensionless Pearson correlation coefficient.}
		\label{fig:EgsCovariance}
	\end{figure}

	We have claimed that the error due to the IMSRG(2) truncation leads to a many-body error which is correlated between neighboring nuclei, thus leading to a greater precision in the separation energies than would be naively estimated.
	To support this claim, we plot in Fig.~\ref{fig:EgsCovariance} the ground-state energy residuals $\delta E=E^{\rm th}-E^{\rm exp}$ for neighboring nuclides.
	We find that the errors in neighboring nuclides are indeed strongly correlated, with a correlation coefficient $\rho_{xy} = \frac{\langle xy\rangle-\langle x\rangle \langle y\rangle}{\sigma_x\sigma_y} \approx 0.9$. (The angle brackets $\langle \, \cdots \rangle$ denote the mean value).
	
	\subsection{Correlation of residuals for a given nuclide}
	
	For a given nuclide, the residuals $\delta S_{n}$ and $\delta S_{2n}$ are correlated, and this should be accounted for when estimating the probability that a given isotope is bound with respect to both $1\mathrm{n}$ and $2\mathrm{n}$ emission.
	This discussion also applies to the proton emission case.
	The proton and neutron separation energy residuals are also correlated (e.g., $\delta S_{n}$ is correlated with $\delta S_{p}$) but this has no impact on the drip line.
	As a result, in the actual calculation, the four-dimensional integral in ~\eqref{eq:pbound} can be factored into a product of two-dimensional integrals.
	The impact of correlations in the residuals reaches at most 0.1 (i.e., changing the probability by 10\%) and is typically below 0.01.
	
	\subsection{Validation}
	
	\begin{figure}[t]
		\centering
		\includegraphics[width=\columnwidth]{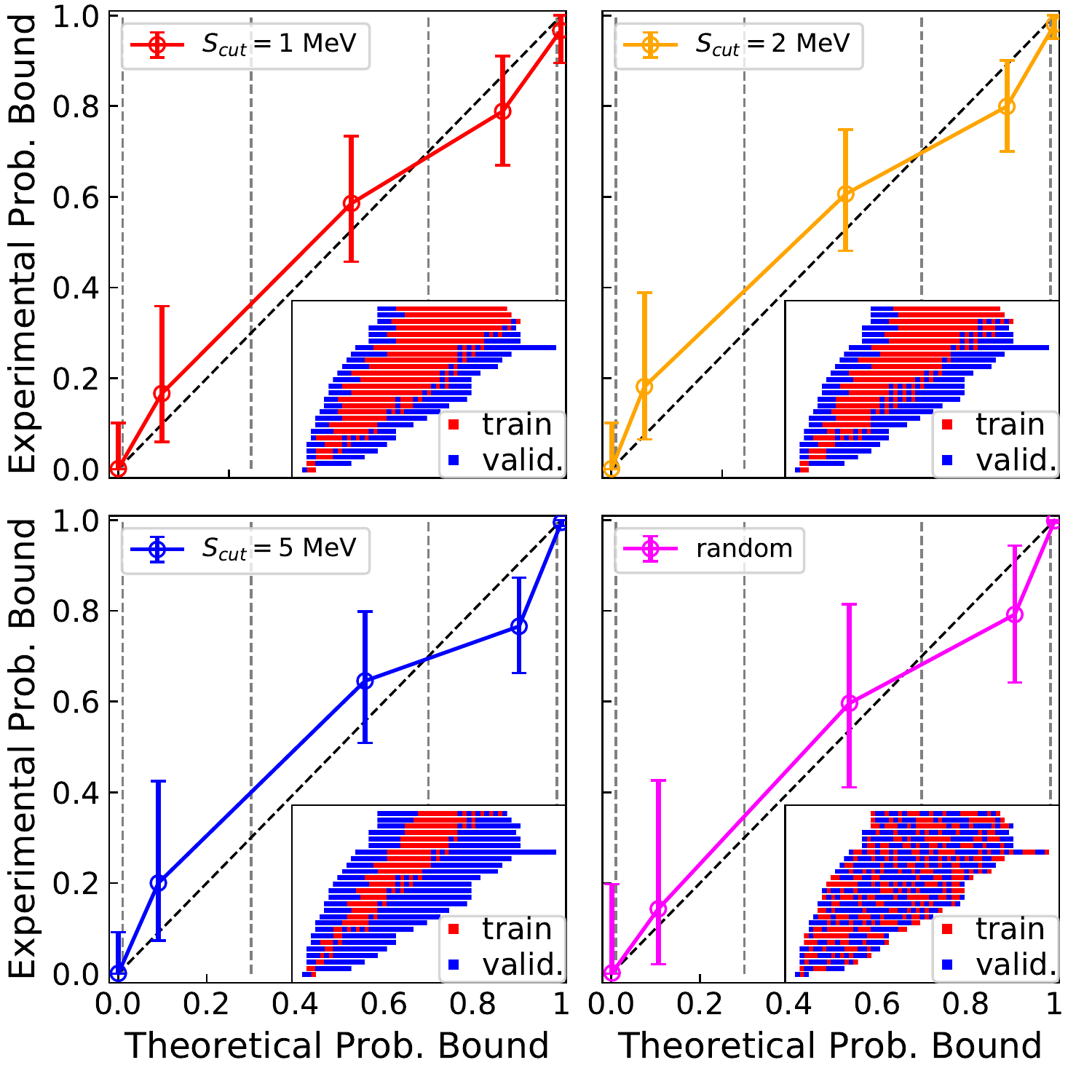}
		\caption{For a given predicted probability to be bound, we plot the corresponding fraction of isotopes that are indeed bound experimentally. A perfect prediction would lie along the dashed line. The error bars represent the 68\% confidence interval estimated using a Wilson score with continuity correction (method 4 from Ref.~\cite{Newcombe1998}). Each panel corresponds to a different partitioning of the data into training and validation sets. This partitioning is illustrated in the inset of each panel, with $N$ on the horizontal axis and $Z$ on the vertical axis.}
		\label{fig:weatherman}
	\end{figure}
	
	To validate the reliability of our approach, we perform an empirical coverage test.
	We partition our data into ``training'' and ``validation'' sets, apply our analysis to the training set, and then bin the validation data according to the predicted probability to be bound.
	In each bin, we evaluate what fraction of the data points are experimentally bound.
	If our predictions are reliable, the fraction bound should coincide with the mean value of the bin.
	
	We perform the partitioning into training and validation sets in two different ways.
	In the first way, we take as training data all nuclides for which all theoretical separation energies are above a cutoff $S_{\rm cut}$, and validate using the remaining cases.
	The first three panels of Fig.~\ref{fig:weatherman} show the results for $S_{\rm cut}=1,2,5$ MeV.
	The second way we randomly partition the data into either set with equal probability.
	The result of this test is shown in the fourth panel of Fig.~\ref{fig:weatherman}, labeled ``random''.
	The vertical error bars represent a 68\% confidence interval, due to finite sample size, estimated using a Wilson score with continuity correction (method 4 from Ref.~\cite{Newcombe1998}).
	The error bars do not get noticeably smaller with increasing $S^{\rm cut}$ because nearly all the additional validation points end up in the upper-right corner with a probability to be bound of approximately 1.
	Importantly, the $S_{\rm cut}=5$~MeV figure demonstrates that we can reliably use well bound nuclei to predict the drip line.
	
	\subsection{Separation energy plots}
	
\begin{figure}[h]
		\centering
		\includegraphics[width=\columnwidth]{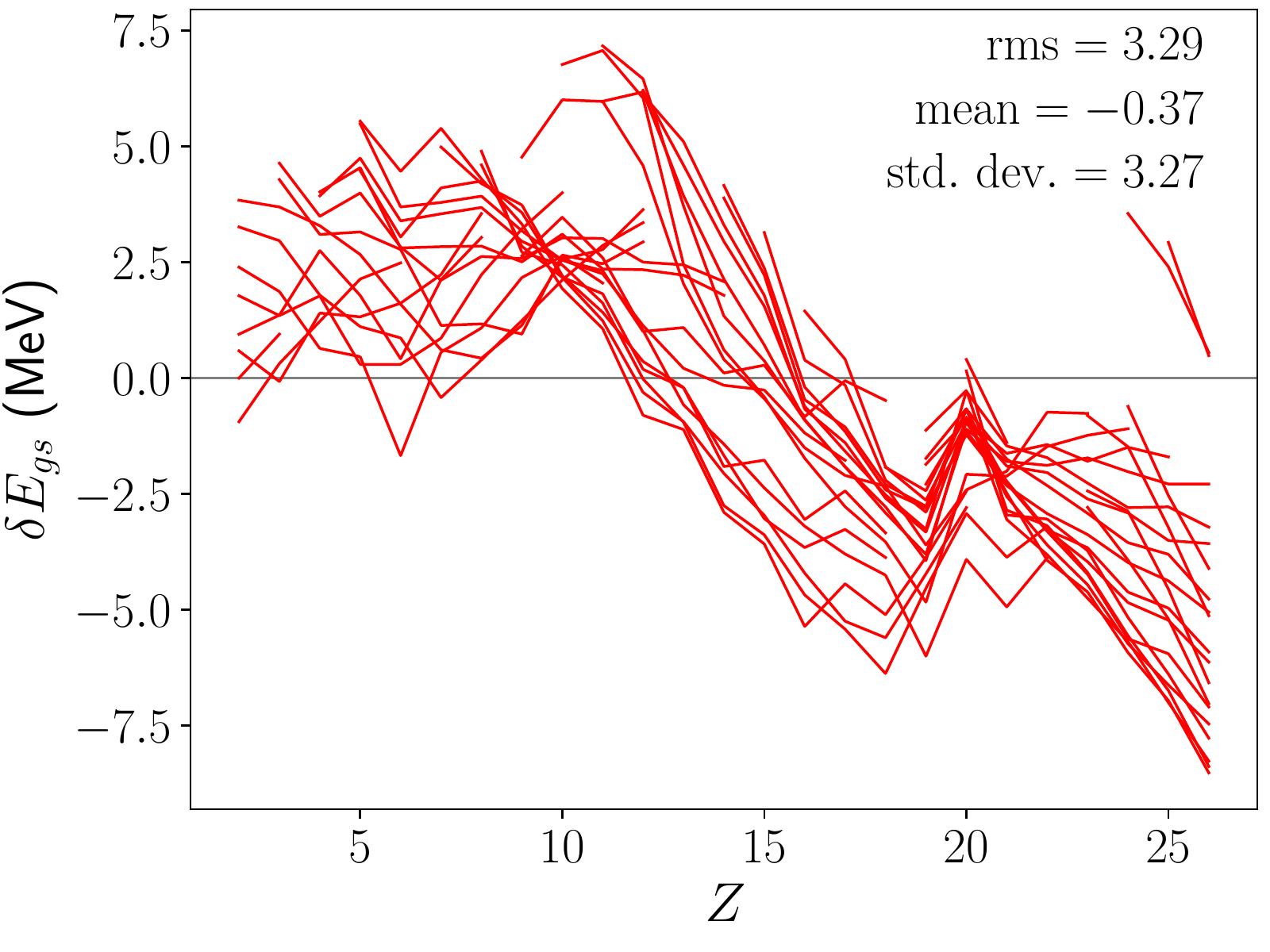}
		\caption{Ground-state energy residuals as a function of $Z$. The different lines connect isotones of constant $N$. The mean, standard deviation, and root-mean-square deviation of the discrepancy are indicated (all in MeV).}
		\label{fig:dEvsZ}
	\end{figure}

	We provide plots of the separation energies, equivalent to Fig.~\ref{fig:cl_sep} in the main text, for all the elements studied here ($2 \leqslant Z \leqslant 26$). In addition, we plot in Fig.~\ref{fig:dEvsZ} the range of agreement with the experimental ground-state energies as a function of $Z$ to complement the information given in the inset to Fig.~\ref{drip_unc} as a function of $N$.
	
	\newpage
	
	\foreach \elemA/\elemB  in {He/Li,Be/B,C/N,O/F,Ne/Na,Mg/Al,Si/P,S/Cl,Ar/K,Ca/Sc,Ti/V,Cr/Mn}{
		\begin{figure*}
			\begin{minipage}{0.98\columnwidth}
				\includegraphics[width=1.0\textwidth]{SupplementPlots/\elemA_5_in_1_column.pdf}
				\caption{Separation energies and probabilities to be bound for the \elemA~isotopes. Blue circles indicate separation energies computed with inconsistent valence spaces.  }
				\label{fig:column_\elemA}
			\end{minipage}\hfill
			\begin{minipage}{0.98\columnwidth}
				\includegraphics[width=1.0\textwidth]{SupplementPlots/\elemB_5_in_1_column.pdf}
				\caption{Separation energies and probabilities to be bound for the \elemB~isotopes. Blue circles indicate separation energies computed with inconsistent valence spaces. }
				\label{fig:column_\elemB}
			\end{minipage}
		\end{figure*}
	}
	
	\foreach \elem in {Fe}{%
		\begin{figure}
			\centering
			\includegraphics[width=0.48\textwidth]{SupplementPlots/\elem_5_in_1_column.pdf}
			\caption{Separation energies and probabilities to be bound for the \elem~isotopes.  Blue circles indicate separation energies computed with inconsistent valence spaces. }
			\label{fig:column_\elem}
		\end{figure}
	}
	
\end{document}